\documentclass[aps,pre,twocolumn,superscriptaddress,showpacs]{revtex4-2}
\usepackage{graphicx}
\usepackage{array}
\usepackage{amsfonts}
\usepackage{amssymb,amsmath,multirow,rotate}
\usepackage{mathrsfs}
\usepackage{booktabs}
\usepackage{threeparttable}
\usepackage{multirow}
\usepackage{epsfig}
\usepackage{threeparttable}
\usepackage{chngpage}
\usepackage{latexsym}
\usepackage{dcolumn}
\usepackage{graphics,graphicx,bm,fleqn,epic,eepic,float}
\usepackage{verbatim}
\usepackage{subfigure}
\usepackage{color}
\usepackage{xr}
\usepackage{float}
\usepackage{listings} 
\usepackage{placeins} 
\usepackage{tikz}
\usepackage{url}
\usepackage{soul}

\definecolor{red}{rgb}{1,0,0}

\definecolor{blue}{rgb}{0,0,1}

\begin{document}

\title{Messaging strategies and the emergence of echo chambers in collective decision-making}

\date{\today}

\author{Ling-Wei Kong} \email{lingwei.kong@cornell.edu}
\affiliation{Department of Computational Biology, Cornell University, Ithaca, NY 14850, USA}

\author{Naomi Ehrich Leonard}
\affiliation{Department of Mechanical and Aerospace Engineering, Princeton University, Princeton, NJ, USA}

\author{Andrew M. Hein} \email{amh433@cornell.edu}
\affiliation{Department of Computational Biology, Cornell University, Ithaca, NY 14850, USA}
\affiliation{Center for Applied Mathematics, Cornell University, Ithaca, NY 14850, USA}

\begin{abstract}
Collective decision-making arises from individual agents integrating their own personal observations with information obtained from social partners. In many biological systems that exhibit collective decision-making, the process by which social information is produced, transmitted, and used is subject to two key constraints. First, individuals often do not observe the internal states or personal observations of their neighbors; instead, they observe neighbors’ discrete actions. Second, agents often have limited attention, such that, at any given moment, only a subset of social partners influences decisions. Using methods from nonlinear dynamics, we show that either of these constraints can cause collective accuracy to become extremely sensitive to the weight individuals place on the information they receive from others. This sensitivity arises from the spontaneous formation of echo chamber–like states in which individuals receive and transmit homogeneous social messages. Under such conditions, collectives become locked in self-reinforcing states that prevent them from tracking changes in the environment. We reveal the mathematical basis of this phenomenon, and show that it emerges not only in generic models of collective decision-making but also in models developed to describe specific biological systems, including neural circuits, eusocial insect colonies, and mobile animal groups. Finally, we identify biologically plausible mechanisms through which individuals may reduce the risk of echo chamber formation and achieve robust yet sensitive collective decisions without requiring fine-tuning parameters. Our results reveal how fundamental constraints on communication shape the dynamics and reliability of collective decisions across diverse biological systems.
\end{abstract}

\maketitle

\section{Introduction}In many biological and social systems, agents decide how to act by combining information gleaned directly from the environment with messages received from others within their social groups. Models of this process have been developed in numerous fields of science including economics \cite{acemoglu2010spread,golub2010naive}, sociology \cite{friedkin1990social}, psychology \cite{sherif1935study,lorenz2011social}, evolutionary biology  \cite{boyd1988culture,whitehead2009evolution}, control theory \cite{leonard2024fast}, animal behavior \cite{torney2015social,kong2025brief,aplin2017conformity}, microbiology \cite{moreno2023quorum}, and neuroscience \cite{klucharev2009reinforcement,basnak2025multimodal}. Despite their similarity, models developed across these domains differ in their assumptions about how agents transform the messages they receive from their neighbors into the messages they send. One common assumption  in such models is that agents directly share either their own personal observations (e.g.,~\cite{galton1907vox,goldstein2014wisdom}) or their own internal states (e.g.,~\cite{degroot1974reaching,jadbabaie2012non}) with other members of their social group. While this assumption may be reasonable in some settings, in many biological and social systems, the messages an agent sends are actually determined by the actions it takes. Actions, in turn, result from internal computations that can quantize internal states, for example if the agent selects from among a set of discrete actions \cite{ceragioli2018consensus}. 

\begin{figure*}
    \centering
    \includegraphics[width=0.95\linewidth]{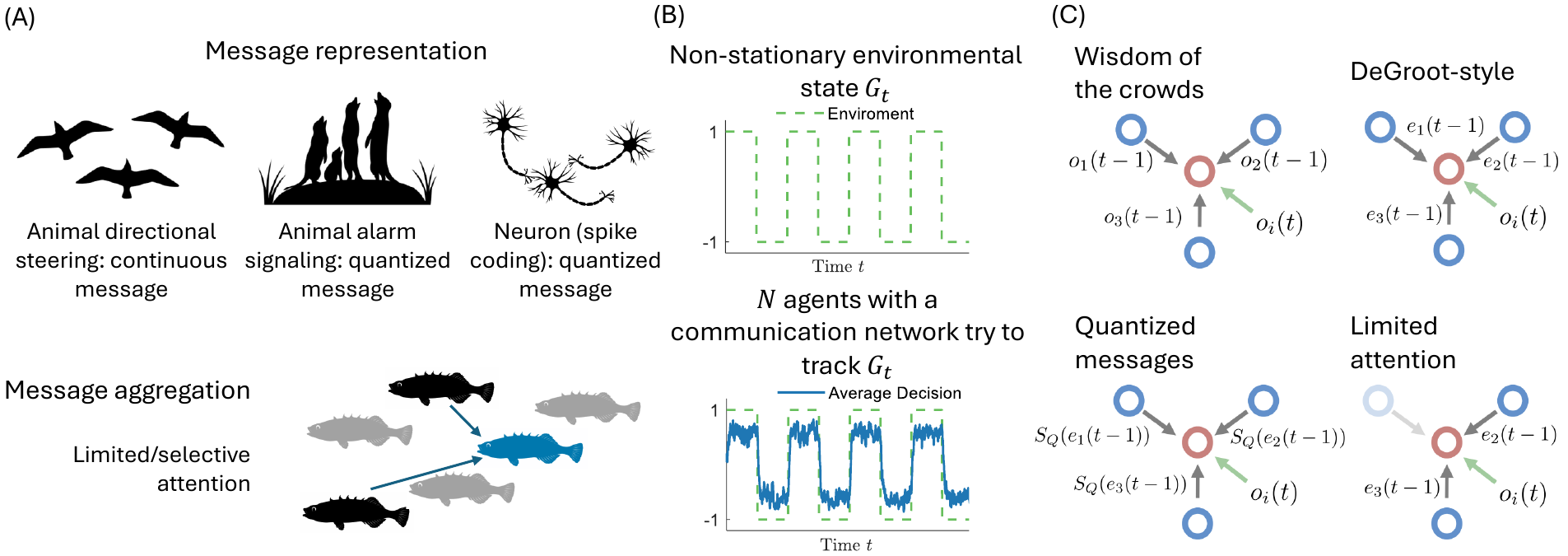}
    \caption{\textbf{Messaging strategies and collective decision-making.}
    (A) Biological systems can differ both in how social messages are represented and in how they are aggregated. Message representations may be continuous, as in animal directional steering, or quantized, as in animal alarm signaling and neuronal spike coding. Social information can also be aggregated selectively, as in limited-attention strategies where only a subset of neighbors strongly influences the focal individual.
    (B) A non-stationary environmental state $G_t$ switches over time and is tracked by a group of communicating agents. Each agent receives a noisy personal observation $o_{i,t}$ and social messages $m_{j,i,t-1}$ from its neighbors, aggregates these inputs to form an internal estimate $e_{i,t}$, and produces both an action $u_{i,t}$ and outgoing social messages.
    (C) The four classes of models studied in this paper differ in their local messaging strategies. In the wisdom of the crowds (WOC) model, agents directly share observations. In the DeGroot-style model, agents share their internal estimates. In the quantized messages model, agents share quantized versions of their internal estimates, which can represent discrete actions. In the limited attention model, agents share estimates as in DeGroot-style models, but only the most salient inputs influence the focal agent's estimate.
    }
    \label{fig:messg_strateg}
\end{figure*}

Examples of biological systems in which discrete actions generate quantized messages  (Fig.~\ref{fig:messg_strateg}A) include animal startle responses \cite{fahimipour2023wild,kong2025brief}, the firing of discrete action potentials leading to release or lack of release of neurotransmitters across chemical synapses (as in ``spike coding'' \cite{rieke1999spikes}), and mate or nest site selection by animals that choose from among a discrete set of options~\cite{giraldeau2002potential}. 

A second widespread feature of decision-making in biological and social systems is limited attention. Although an agent may receive messages from many social partners, it often bases its decisions on only a subset of these inputs. Limited attention is widespread in human and animal decision-making, ~\cite{greenfield2021rhythm,lemasson2009collective,van2010selective,puy2024selective}. In some cases, it reflects cognitive or sensory constraints, as exemplified by the classic ``cocktail party problem'' in which messages from multiple sources conflict with one another, requiring focus on a single source \cite{mcdermott2009cocktail}. In other cases, agents may selectively prioritize the most salient or informative signals, placing greater weight on a subset of social messages \cite{lemasson2009collective,lemasson2018motion}. In either case, limited attention effectively restricts the information that is incorporated into decision-making, introducing an additional constraint on collective information processing.

While quantized messages and limited attention are widespread in biological systems, their effects on collective decision-making dynamics remain poorly understood. 
Past work on engineered multi-agent systems has shown that communication constraints such as quantization or limited channel capacity can degrade the accuracy or convergence rate of distributed algorithms~\cite{msechu2008decentralized,nedic2009distributed,li2010distributed}. 
Recent work in opinion dynamics further demonstrated that quantized communication can induce nonconsensus equilibria and large deviations from consensus~\cite{ceragioli2018consensus}. However, these effects depend on network topology, are diminished on highly mixed graphs, and primarily reflect the group's failure to reach agreement. 
Related work has also shown that the coupling between continuous internal opinions and discrete expressed choices can generate qualitatively distinct collective outcomes, such as polarization and pluralistic ignorance~\cite{aghbolagh2023coevolutionary}.

Here, we ask whether message quantization and limited attention merely degrade collective performance, or whether they can fundamentally alter the nature of collective decision-making.
In particular, we show that when agents need to track a changing environment, these constraints can give rise to echo chambers with hysteresis or lock-in that persists even on fully connected networks. This suggests a qualitatively different failure mode from previously studied topology-dependent nonconsensus effects~\cite{ceragioli2018consensus}  and from the collective outcomes studied in opinion--action coevolution models~\cite{aghbolagh2023coevolutionary}.

We consider four classes of models that originate from distinct fields -- psychology, economics, collective animal behavior, and neurobiology -- but all pertain to the problem of how agents within collectives integrate personal observations and social messages to make decisions (Fig. \ref{fig:messg_strateg}C). These models can be cast within a unified framework by noting that the principal difference among models is their assumptions about how agents transform their observations of the environment, and the messages they receive, into the messages they send (Fig. \ref{fig:messg_strateg}A-B).

In one class of models, messages are simply agents' own personal observations. In a second, agents directly communicate their own internal states. In the third, messages are some discretization of agents' internal states, leading to a quantized set of possible messages. And in the fourth, agents can only attend to a limited number of sources at once.  We ask how these distinct assumptions, which correspond to different classes of messaging strategy influence dynamics and performance of collective decision-making in a dynamic environment.

\section*{Models of collective decision-making}

In the four classes of models studied here, the decision made by each agent at time, $t$, is a function of that agent's estimate of the true state of the environment. In all four models, estimates have the general form 

\begin{align}
    e_{i,t} = \frac{1}{1+k_i\omega}\left( o_{i,t} + \omega\sum_{j} A_{ij}\eta_{ij,t} m_{j,t-1}\right).
      \label{eq:gen_model_e}
\end{align}
Here, $o_{i,t}$ is the personal observation of agent $i$ at time $t$, $m_{j,t-1}$ is the message sent at time $t-1$ by the $j$th neighbor of agent $i$, $\omega$ is the social weight agents apply to messages from neighbors, $k_i$ is the in-degree of agent $i$,  and $A\in \{0,1\}^{N\times N}$ denotes the adjacency matrix of the network of communication where, $A_{ij}=1$ if agent $i$ receives messages from agent $j$. The variable $\eta_{ij,t}$ denotes an attention mask that determines whether agent $i$ pays attention to agent $j$ at time $t$. In most model variants $ \eta_{ij,t}$ is always equal to one for all $i,j$. The exception is that in the limited attention model (described below), $\eta_{ij,t}$ can be equal to zero or one.  Here and throughout, we use the term ``message'' in a very general sense, to include deliberately shared information or signals as well as cues produced when one agent takes an action that can be perceived by other agents (e.g., visual or auditory cues produced during an escape response, production of metabolic byproducts).

Under the general formulation of \eqref{eq:gen_model_e}, social weight, $\omega$, can be interpreted as the ratio of the weight applied to social messages and the weight applied to personal observations. For instance, a value of $\omega = 0.5$ means that agents weigh their personal observations twice as strongly as they weigh the social messages they receive. In the following sections, we describe how the four model classes relate to this overarching model formulation.

\subsection*{Naive wisdom of the crowds (WOC): 
 agents share their personal observations} One of the earliest collective decision-making models assumes that agents simply share their raw personal observations and make decisions based on 
 the public pool of observations~(\cite{galton1907vox}, Fig.~\ref{fig:messg_strateg}C). 
Following Refs.~\cite{galton1907vox,surowiecki2005wisdom,goldstein2014wisdom,lorenz2011social}, we consider an observation-sharing model in which agents transmit their own personal observations to neighbors, rather than their personal states or beliefs. Dynamics of estimates of the environment are given by \eqref{eq:gen_model_e} with $\eta_{ij,t} = 1$ and 
\begin{align}
    m_{i,t}=o_{i,t}.
\end{align}

\subsection*{DeGroot-style model: agents share their internal states}

In seminal work on consensus formation, DeGroot~\cite{degroot1974reaching} developed a model in which agents iteratively update their beliefs by averaging those of their neighbors. Many subsequent studies have explored implications of this framework in various settings 
(e.g.,\cite{friedkin1990social,demarzo2003persuasion,olfati2004consensus,golub2010naive}). 
Braca et al.~\cite{braca2008enforcing} introduced the ``running consensus'' model,  in which agents continue to collect new observations of an environmental signal while simultaneously averaging with their neighbors, thereby extending DeGroot-style social averaging to an online distributed estimation setting closely related to ours.
In a similar spirit, Olfati-Saber~\cite{olfati2005distributed} showed that distributed Kalman filtering can be constructed by combining local micro-Kalman filters with embedded consensus filters, illustrating how consensus-based information exchange can support decentralized estimation of a dynamic process from noisy measurements.
Related work has also examined how network structure and the placement of leaders or directly informed nodes shape coherence and tracking error in noisy leader--follower networks, including Patterson and Bamieh~\cite{patterson2010leader} and Fitch and Leonard~\cite{fitch2015joint}.
Jadbabaie et al.~\cite{jadbabaie2012non} studied a related repeated-observation setting, but in a non-Bayesian social learning framework where each new private signal is first incorporated through an individual Bayesian update before being socially combined with neighbors’ beliefs. This effectively introduces agent-level memory. In contrast, we initially assume agents have no individual memory (we explore effects of individual memory in SI Section~\ref{sec:SI_individualmemory}).

In the DeGroot-style model, the social message sent by agent $i$ in each timestep is simply its current estimate of the environmental state (Fig. \ref{fig:messg_strateg}C). Estimates of the state of the environment are given by \eqref{eq:gen_model_e} with $\eta_{ij,t} = 1$ and 
\begin{align}
m_{i,t}=e_{i,t}.\label{eq:deg_m}
\end{align}

\subsection*{Quantized message model: 
 agents share quantized versions of their estimates}

Torney et al.~\cite{torney2015social} developed a collective decision-making model that uses only the actions taken by agents as social messages, and applies a Bayesian rule to aggregate personal observations and social messages to make decisions. We incorporate this model into our framework, and (in SI Section~\ref{sec:SI_Torney}) show that, with small modifications, this model is equivalent to a DeGroot-style model with quantized messages (Fig. \ref{fig:messg_strateg}C). Dynamics of estimates of the environmental state are given by \eqref{eq:gen_model_e} with $\eta_{ij,t} = 1$ and 
\begin{align}
    m_{i,t}=f_Q(e_{i,t}),
\end{align}
where $Q$ is a discrete set of possible messages, and function $f_Q(\cdot)$ maps the continuous estimate to the closest possible discrete message.

\subsection*{Limited attention model: agents share internal states but some sources are ignored}

In many biological systems, agents only attend to a limited number of sources at a time. This phenomenon is sometimes called limited attention or selective attention, because an agent is only influenced by stimuli from a limited number of sources \cite{mcdermott2009cocktail}, or in some cases, a single source~\cite{broadbent2013perception}. 
Which sources influence decision-making and which are ignored often depends on the salience of messages from each source. For example, there is evidence that agents often attend to the neighbors that produce the strongest stimuli (e.g., \cite{lemasson2009collective,lemasson2018motion}). 

Since \eqref{eq:gen_model_e} assumes messages are real numbers (i.e., $m \in \mathbb{R}$), a natural way to quantify the strength of a given source is as the magnitude of the message received. We therefore consider a limited-attention variant in which, at each timestep, an agent updates its estimate using only the $L$ strongest messages it receives. The dynamics are still governed by \eqref{eq:gen_model_e} and \eqref{eq:deg_m}, except that $\eta_{ij,t} = 1$ only for the $L$ sources whose messages have the largest absolute values at that timestep (Fig. \ref{fig:messg_strateg}C; see SI Section~\ref{sec:SI_TopOM}, for alternative scenario in which personal observations compete for salience with social messages).
Sources that are ignored in a given timestep $t$ have $\eta_{ij,t} =0$.
In the space of collective decision-making models, this limited attention model can be thought of as part of a broader class of soft/hard winner-take-all models~\cite{maass2000computational,feldman1982connectionist}.

\subsection*{Environmental dynamics and task structure} We study each model in a canonical collective decision-making task in which agents must select actions that match a changing latent environmental state \cite{torney2015social,boyd2013evolutionary, whitehead2009evolution, kong2025brief}. This state may represent, for example, the presence or absence of danger, the location of resources, or other time-varying environmental features. Following previous studies, we adopt a discrete-time formulation~\cite{torney2015social,jadbabaie2012non}. In each timestep, each agent tries to choose the action that is best matched to the true (latent) state of the environment at that time. The payoff for an agent is determined by its average action accuracy over a time scale encompassing many changes in environmental state. At each timestep $t$, the environment is characterized by a binary state $G_t\in\{-1,+1\}$. In the main text we consider periodic switching, where $G_t$ flips every $T$ time steps (results with stochastic switching are presented in SI Section~\ref{sec:SI_EnvNoise}). Each of the $N$ agents seeks to infer $G_t$ while exchanging messages through a communication network (Fig.~1A), making the task a collective decision problem.

Each agent $i$ receives a noisy personal observation of the environmental state,
\begin{align}
    o_{i,t}=G_t+\sigma_n\zeta_{i,t}, \label{eq:o}
\end{align}
where $\sigma_n$ controls the noise magnitude and $\zeta_{i,t}$ represents observational noise that may be temporally autocorrelated. Following Ref.~\cite{torney2015social}, we assume temporal autocorrelation within each agent's observations but no correlation across agents (alternative assumptions are discussed in SI Section~\ref{sec:SI_EnvNoise}).

At each time step, based on its estimate $e_{i,t}$, agent $i$ selects a binary action $u_{i,t}$ representing its best guess of the true environmental state. The payoff of each agent is defined as the average accuracy of its actions over time. The estimate is mapped to an action according to
\begin{align}
    u_{i,t} = \text{sign}(e_{i,t}). \label{eq:action}
\end{align}

\subsection*{Social network structure} 
The structure of communication networks can influence the effect and outcome of social interactions (e.g., \cite{tarnita2009strategy}). Here, we focus on populations with structured communication networks. More specifically, we assume that, at time $t$, agent $i$ broadcasts the same message $m_{i,t}$ to all its neighbors over the directed communication network defined by the adjacency matrix $A$. The in-degree of agent $i$ is $k_i=\sum_j A_{ij}$, which quantifies the number of neighbors sending messages to agent $i$. We denote the average in-degree of all the agents as $\langle k \rangle=\sum_ik_i/N$. This parameter, $\langle k \rangle$, can be used to tune the connectivity of agents within the group.

We mainly focus on a canonical network structure, the Erdős–Rényi (ER) random network~\cite{erd6s1960evolution,newman2018networks} (Fig.~\ref{fig:messg_strateg}D). In this topology, connections between pairs of agents are formed independently with a fixed probability, producing a network that is statistically homogeneous on average while still allowing fluctuations in local connectivity. We also study four other representative network topologies with different structural properties in the main text. Additional analyses, including seven more network types as well as variations in network size $N$ and average degree $\langle k \rangle$, are presented in SI Sections ~\ref{sec:SI_heter} and ~\ref{sec:SI_addnets}.

\section*{Results}

\begin{figure*}[htb!]
    \centering
    \includegraphics[width=0.465\linewidth]{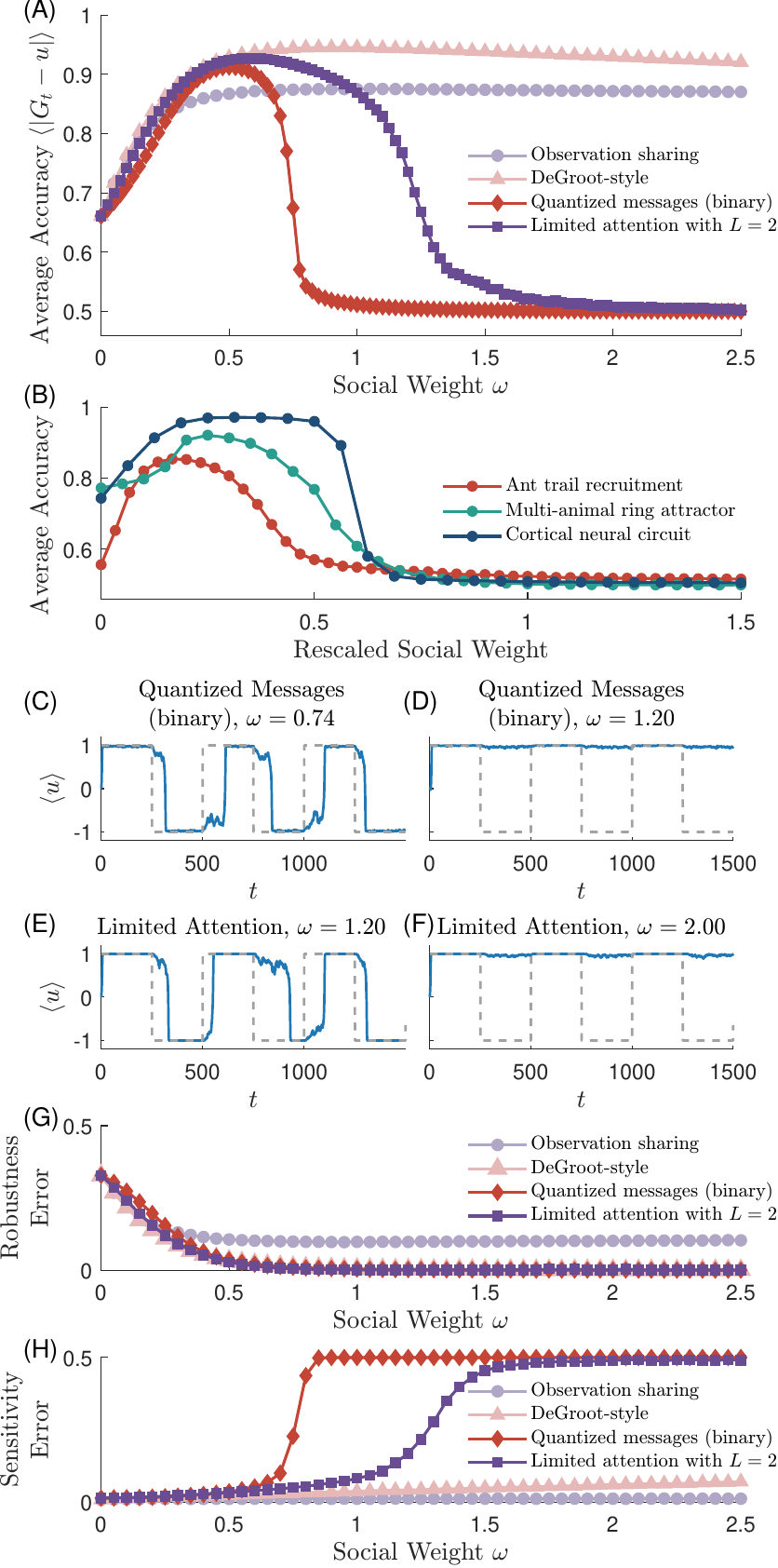}
    \caption{
    {
    \textbf{Dependence of collective accuracy on social weight.}
    \textbf{(A)} Average decision accuracy as a function of social weight,  $\omega$. The quantized message model and the limited attention model exhibit a sharp collapse in accuracy beyond a critical $\omega$. In contrast, the models without bottlenecks do not exhibit such abrupt transitions to low-accuracy regimes.
    \textbf{(B)} A similar sudden drop in average accuracy at high social weight is also observed across three models with greater biological grounding.
    \textbf{(C, D)} Time series of the average decision $\langle u \rangle$ in the in the quantized message (binary) model for $\omega$ near and beyond the transition. In (C), as $\omega$ approaches the critical value, the group exhibits an increasingly delayed response to changes in the environment. In (D), beyond the transition, the group becomes effectively unresponsive, entering a locked-in state that persists despite subsequent environmental switches.
    \textbf{(E, F)} Time series of the average decisions $\langle u \rangle$ in the limited attention model for $\omega$ near and beyond the transition, showing qualitatively similar behavior to (C,D): delayed tracking near the transition (E) and lock-in with near-complete loss of responsiveness beyond it (F).}
    \textbf{(G, H)} Decomposition of decision errors into (G) robustness and (H) sensitivity errors. Results are shown for four model classes: an observation-sharing model, a DeGroot-style model, a quantized message model with binary messages, and a limited attention model. Across all four models, robustness error decreases monotonically with increasing social weight $\omega$, while sensitivity error increases monotonically. In the quantized message model, sensitivity error can exhibit an abrupt increase before reaching an error of 0.5, equivalent to completely random decisions. The sensitivity error of the selective attention model also increases to 0.5. Unless otherwise noted, simulations in all figures were conducted on networks with $N = 512$ agents and $\langle k \rangle = 8$. Robustness to system size and additional results for other average in-degrees are provided in SI Section~\ref{sec:SI_addnets}.}
    \label{fig:sw}
\end{figure*}

Given the noise in personal observations of the environment (\eqref{eq:o}), agents inevitably make errors when seeking to match their actions to the state of the environment \cite{kong2025brief}. However, the frequency and nature of these errors depends strongly on the social weight, $\omega$.  Fig.~\ref{fig:sw} shows results from the four model classes. When social weight $\omega=0$, agents in all four models make decisions based solely on their own noisy personal observations. In all four models, as $\omega$ increases from zero, the average accuracy, computed across agents and over time, also increases. 

A second feature shared by all models is that, beyond an intermediate range of social weight $\omega$, further increasing social influence begins to reduce average accuracy. In the observation sharing and DeGroot-style models, this reduction is gradual. By contrast, the quantized message and limited attention models show an abrupt transition at intermediate $\omega$, beyond which mean accuracy drops sharply (Fig.~\ref{fig:sw}A), eventually falling below the performance achieved when agents ignore social information altogether ($\omega = 0$).

As Fig.~\ref{fig:sw}A shows suboptimal performance at both low and high social weights $\omega$, we asked whether these failures arise from the same mechanism or from distinct ones. While the reduced accuracy at low $\omega$ is expected, as agents rely primarily on noisy personal observations, the decline in performance at high $\omega$ suggests a qualitatively different mechanism.
To investigate this, we examined the temporal dynamics of population-averaged accuracy as the environment changes (Fig.~\ref{fig:sw}C--F). At large $\omega$, the collective exhibits substantial delays between environmental flips and changes in the majority decision, despite occasional periods of accurate tracking (Fig.~\ref{fig:sw}C and E). As $\omega$ increases further, these delays become more pronounced, and in the extreme large-$\omega$ regime, the average decision of the group becomes effectively insensitive to the environment (Fig.~\ref{fig:sw}D and F). In this regime, the majority action no longer responds to environmental changes, a pattern also emphasized by \cite{torney2015social}.

The breakdown of collective tracking in the high-social-weight regime is characterized by three related but distinct features. Strong reciprocal social influence creates \emph{echo chamber dynamics}. This, in turn, produces \emph{collective distortion}, in the sense that socially transmitted messages no longer faithfully represent the environmental state. The resulting group-level dynamical outcome is \emph{lock-in}, with delayed or absent responses to environmental change.

\subsection*{Decomposition of agents' errors}

We next distinguish these two forms of suboptimal behavior in terms of the dominant source of error: \emph{robustness error} and \emph{sensitivity error}. Robustness error, which dominates in the low-$\omega$ regime, refers to incorrect decisions arising because each agent receives only a noisy personal observation of the environment. Sensitivity error, by contrast, arises from delayed or absent responses to environmental changes, reflecting insufficient sensitivity to new observations. Fig.~\ref{fig:sw}G--H shows how these two types of error vary with social weight $\omega$ in the four models. In all models, robustness error decreases monotonically with increasing $\omega$, whereas sensitivity error increases monotonically. The tradeoff between these two sources of error produces a unimodal relationship between average accuracy and social weight $\omega$.

The robustness error results from the noise inherent in agents' personal observations of the environment, and this error is reduced as individuals rely more heavily on social messages to make decisions. This noise buffering effect of using socially-transmitted messages during collective decision making has been studied extensively elsewhere ~\cite{kao2014decision,goldstein2014wisdom,winklmayr2023collective}. 
However, why the sensitivity error grows so rapidly in quantized message and limited attention models as $\omega$ increases beyond its optimal value is less clear.

\subsection*{To exploit benefits of collective decision-making while avoiding costs, social weight must be tuned to statistics of both the network and the environment}
One pattern that is clear in Figure~\ref{fig:sw}A is that quantized message models and limited attention models can exhibit high collective accuracy, but only if social weights of agents are tuned within a fairly narrow range. Moreover, the fact that accuracy declines precipitously for social weights that are higher than the optimal values implies that mis-tuning of social weights could be extremely costly. In SI, we show that the optimal social weight and the value of social weight at which performance drastically decreases depends on the properties of network that defines who sends messages to whom (Fig. ~S\ref{fig:SI_heterNets}) and temporal dynamics of the environment (Fig. ~S\ref{fig:addEnv}). An implication is that, in order to maintain high performance, individuals within a collective would need a way to tune social weights in response to both changes in network topology and changes in environmental dynamics. How such a coordinated, dynamic parameter tuning could be achieved is not clear.

\subsection*{
Models with quantized messages or limited attention exhibit multistability when social weight is high}
\begin{figure*}[htb!]
    \centering
    \includegraphics[width=0.45\linewidth]{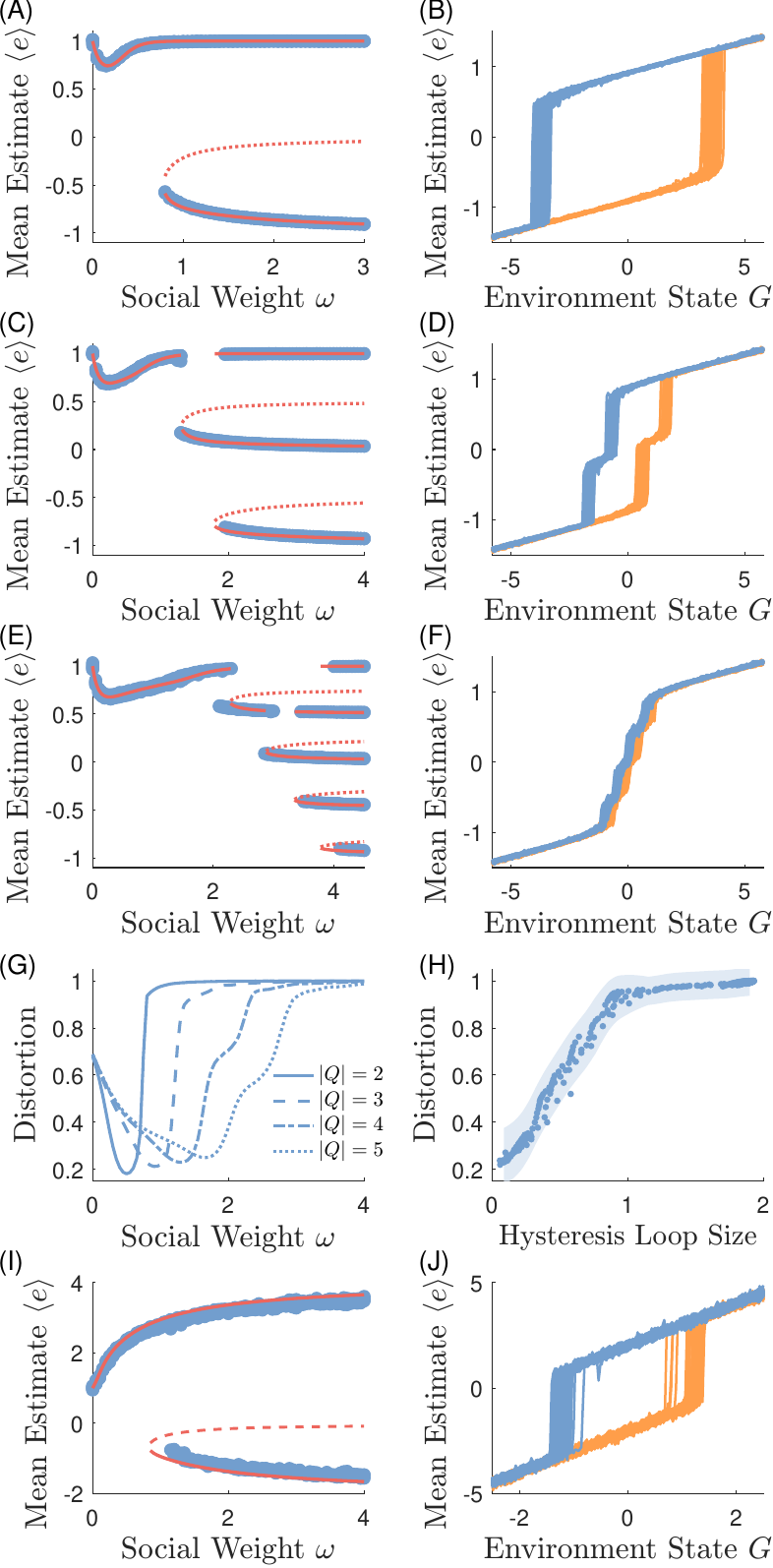}
    \caption{
    \textbf{Coexistence of multiple steady states and hysteresis in collective decision-making.}
    \textbf{(A, C, E)} Coexistence of multiple steady states of $\langle e \rangle$ in quantized message models with (A) binary messages, (C) $Q=\{-1,0,+1\}$, and (E) $Q=\{-1,-0.5,0,+0.5,+1\}$. The blue dots are results from simulations of the full agent-based models. The red solid curves denote the reachable stable branches of fixed points from the mean-field theory. The red dashed curves denote the branches of the unstable fixed points.
    \textbf{(B, D, F)} Hysteresis loops in the three models as in panels (A, C, E), respectively. While we assume the environment state to be binary $G=\pm1$ in the models, for demonstration of the hysteresis loops, we allow $G$ to take a wider range of values. We gradually sweep $G$ from a high value to a low value (blue branches) or sweep $G$ from a low value to a high value (orange branches), and record the time series of mean estimate $\langle e \rangle$ over $N=512$ agents in a group. We run 50 independent trials for each branch. As the messages are quantized with finer steps, the hysteresis loop shrinks.
    \textbf{(G)} Distortion between social messages $m$ and the environment state $G$ with different quantization resolution. The distortion is calculated by the average of $|m-G|$.
    \textbf{(H)} A wider hysteresis loop usually leads to higher distortion. The dots are collected with different quantization resolution $|Q|\in \{2,3,4,5\}$ and different social weight $\omega\in[0.1,4]$.
    \textbf{(I)} Coexistence of multiple steady states in the limited attention model with $L=2$.
    \textbf{(J)}  Hysteresis loops in the limited attention model. 
    }
    \label{fig:multi}
\end{figure*}

One explanation for the abrupt drop in performance in the quantized message and limited attention models when $\omega$ exceeds its optimal value is a change in qualitative features of collective decision-making in this regime. To test this, we performed simulations in which the environment was constant, and computed agents' mean estimate of environmental state across a range of social weight values $\omega$. We found that in both the quantized message model (Fig. \ref{fig:multi}(A, C, E)) and limited attention model (Fig.~\ref{fig:multi}(I)), there is one or multiple critical values of $\omega$ beyond which the number of possible stable states of the system increases. Above this critical value, agents' estimates of the environment may converge to distinct values, even when the environment is not changing, and this pattern implies a bifurcation from a single equilibrium to multiple equilibria.

In contrast to the patterns shown in the quantized message and limited attention models, the observation sharing and DeGroot-style models exhibit a single stable equilibrium regardless of social weight $\omega$ (SI Section~\ref{sec:SI_mf}).

\subsection*{Mean-field theory: reduced iterative map and its bifurcation structure}

To better understand the origin of multistability and hysteresis, we develop a simplified mean-field model that reduces the collective dynamics to a two-dimensional iterative map. The red solid curves in Fig.~\ref{fig:multi} (panels A, C, E, I) show the mean-field results of the stable fixed points, which agree well with results of the full simulations (Fig.~\ref{fig:multi}A, C, E, I blue points). 

To construct the mean field approximation, we assume that the internal estimates $\{e_{i,t}\}$ are i.i.d.~Gaussian random variables with mean $\mu_t$ and variance $\sigma_t^2$. We also assume the observational noise is i.i.d. Under these assumptions, the quantized message model yields the iterative map
\begin{align}
    \mu_t &= (1-\omega_s) G_t + \omega_s M_{t-1}^{(1)} \label{eq:map_mu}\\
    \sigma_t^2 &= (1-\omega_s)^2 \sigma_n^2 + \omega_s^2 \frac{M_{t-1}^{(2)}}{k},    \label{eq:map_sigma}
\end{align}
where $M_{t-1}^{(1)} = \mathbb{E}(m_{i,t-1})$ and $M_{t-1}^{(2)} = \mathrm{Var}(m_{i,t-1})$ are the mean and variance of social messages, $\sigma_n$ is the standard deviation of the observational noises, and $\omega_s=1-1/(k \omega +1)$ is the total relative weight of social messages. When the network in-degree is not uniform (e.g., in ER networks), we replace $k$ in Eq.~\ref{eq:map_sigma} by $k^*=1/\langle 1/k \rangle$.
For binary messages $Q=\{-1,+1\}$, the map reduces to
\begin{align}
\mu_t &= (1-\omega_s) G_t + \omega_s [2 \Phi(\frac{\mu_{t-1}}{\sigma_{t-1}}) -1],\\
\sigma_t^2 &= (1-\omega_s)^2 \sigma_n^2 
+ \frac{\omega_s^2}{k} \{1- [2 \Phi(\frac{\mu_{t-1}}{\sigma_{t-1}}) -1]^2\},
\end{align}
where $\Phi$ is the cumulative distribution function of a standard Gaussian distribution. The explicit forms of Eqs.~\ref{eq:map_mu}-\ref{eq:map_sigma} for a generic quantization set $Q$ are shown in SI Section~\ref{sec:SI_mf}.

Under similar assumptions, the limited attention model also yields a closed two-dimensional map (see SI Section~\ref{sec:SI_mf} for details). Approximating top-$L$ selection by a two-sided truncation, we obtain
\begin{align}
\mu_t &= \frac{G_t + L\,\omega\,\mu_{\mathrm{sel},t-1}}{1+L\,\omega}, \\
\sigma_t^2 &= \frac{\sigma_n^2 + L\,\omega^2\,\sigma^2_{\mathrm{sel},t-1}}{(1+L\,\omega)^2},
\end{align}
where $\mu_{\mathrm{sel}}$ and $\sigma^2_{\mathrm{sel}}$ are the conditional mean and variance of $m$ given $|m|>q$).

Using the reduced mean-field maps, we analyze the bifurcation structure of both the quantized message and limited attention models. We find that the observed changes in the number and accessibility of stable fixed points arise from two distinct mechanisms.
The first mechanism is the catastrophic bifurcations leading to the emergence of additional stable branches as the social weight $\omega$ increases. In Fig.~\ref{fig:multi} (A) and (I), these transitions take the form of unfolded supercritical pitchfork bifurcations: when $G_t=0$, they reduce to standard supercritical pitchfork bifurcations, whereas for $G_t\neq 0$, explicit symmetry breaking unfolds the pitchfork. Interestingly, this class of bifurcations has been intensively studied in the context of a recently developed continuous model of decision-making known as nonlinear opinion dynamics \cite{leonard2024fast}.
In Fig.~\ref{fig:multi} (C) and (E), where the message set contains more quantization levels, the corresponding transitions appear as a sequence of fold bifurcations. More generally, these transitions can be understood as slices through a higher-dimensional bifurcation structure in the joint parameter space of $(\omega, G)$. In SI section~\ref{sec:SI_mf}, we show that this organization is consistent with an underlying cusp bifurcation by visualizing the two-parameter bifurcation structure and analyzing the eigenstructure of the mean-field maps near the transition points.

The second mechanism does not correspond to a bifurcation of the map itself. Instead, it arises from changes in the reachability of stable fixed points in the switched dynamical system as the environment changes. In our model, the environment switches over time between $G_t=+1$ and $G_t=-1$, such that both the full model and the reduced map should be viewed as switched systems composed of two subsystems, one for each value of $G_t$. Although each subsystem may possess its own stable fixed points, not all of these fixed points are dynamically reachable in the full switched system 
because the initial condition for one subsystem is constrained to be the final state of the other subsystem, which can be outside the basin of attraction of certain stable fixed points of the former subsystem.

\subsection*{In quantized message and limited attention models, multistability leads to collective lock-in}

The existence of multistability in quantized message and limited attention models need not necessarily lead to a decrease in performance. 
However, the symmetry of the system dictates that when new stable fixed points appear, they tend to lie near collective states that correspond to the *wrong* environmental state. Specifically, when $G_t = +1$, the bifurcation creates an additional stable fixed point near $\langle e \rangle \approx -1$, and vice versa. In a binary switching environment, this symmetry has a direct consequence: when the environment flips from $G_t = +1$ to $G_t = -1$, the collective state that was previously tracking the correct environmental state now falls within the basin of attraction of the wrong fixed point under the new environment. The collective therefore remains locked in a state that no longer matches the environment, producing the lock-in dynamics observed in Fig.~\ref{fig:sw}(D, F). Even before the bifurcation occurs — when $\omega$ is slightly below the critical value — the remnant of the incipient fixed point (a so-called ghost state) creates a region of slow dynamics in state space, causing substantial delays in the collective's response to environmental switches (Fig.~\ref{fig:sw}(C, E)).

A complementary perspective on this phenomenon emerges when the environmental state $G$ is treated as a continuous parameter rather than a binary variable. In this view, the drop in performance is a consequence of hysteresis.
Hysteresis refers to history dependence, whereby the state of a system depends not only on the current conditions but also on the path by which those conditions were reached~\cite{strogatz2024nonlinear}. In such systems, the transition between alternative states occurs at different parameter values depending on whether conditions are increasing or decreasing, forming a hysteresis loop. Figure \ref{fig:multi} (B, D, F) shows the hysteresis loops in models with three different quantized message sets, and Fig.~\ref{fig:multi} (J) shows the same curves for the limited attention model. 
Because the hysteresis loops can extend well beyond $G_t = \pm 1$, environmental switches of this magnitude are insufficient to push the collective across the transition, producing the lock-in described above.

In terms of the microscopic dynamics of interactions among agents, the bifurcations and corresponding hysteresis loops in Fig.~\ref{fig:multi} result from echo chamber-like states within the collective; when social weight becomes large, an agent can become locked in suboptimal states when the environment changes because of reciprocal influence with its neighbors. The social messages agents send are a function of agents' estimates of the true state of the environment, but because of hysteresis, agents may reach the same estimates under very distinct states of the environment. We refer to this phenomenon as collective distortion \cite{kong2025brief} because reciprocal feedback among agents distorts the information about the true state of the environment that agents glean through their own personal observations. 
Figure~\ref{fig:multi}(G,H) makes this connection explicit. We define distortion in the standard sense \cite{Cover2009} as the mean error between the true environmental state $G$ and the message $m$ produced by an agent. Distortion is minimized at intermediate values of the social weight $\omega$, across different levels of message quantization (Fig.~\ref{fig:multi}G), and the region of low distortion becomes broader as the message set becomes richer. 
As shown in Fig.~\ref{fig:multi}(H), distortion increases with the size of the hysteresis loop, indicating that collective memory effects directly degrade the fidelity of information transmission. 
Increasing the resolution of message quantization reduces the hysteresis loop size (Fig.~\ref{fig:multi}B,D,F), thereby lowering distortion at a given $\omega$. Here, the hysteresis loop size is measured as $\max\{|\langle e\rangle - G|\}$ at $G=\pm1$ on the loop.

\subsection*{Two routes to lock-in}

To analyze the mechanism of social interactions in a more general way, we can decompose the system into an \emph{encoding function} $f$ mapping estimates to outgoing messages, and an \emph{aggregation function} $h$ combining incoming messages:
\begin{align}
    e_{i,t} = h \circ f (\{e_{j,t-1}\}), \quad j\in\{j\,|\,A_{ij}=1\}. \label{eq:general_fg}
\end{align}
The two models studied above, the quantized message and limited attention models, illustrate two distinct routes by which nonlinearities generate additional stable fixed points. In the quantized-message model, $h$ is linear averaging and the nonlinearity resides in $f$: the stable consensus fixed points are the stable fixed points of $f$, which in the limit $\omega\to\infty$ correspond to the quantization levels $Q$. In the limited attention model, $f$ is the identity and $h$ averages the top-$L$ messages by magnitude. In this case, every perfect consensus is a fixed point and the consensus-level analysis is degenerate. Instead, the nonzero steady states arise from a distributional mechanism: when agents' estimates have nonzero variance, the top-$L$ selection by absolute value preferentially samples messages aligned with the mean, effectively amplifying it. This amplification 
sustains an isolated steady state at the population level. A detailed analysis of both cases is provided in SI~Section~\ref{sec:SI_consensus}.

\subsection*{Multistability and hysteresis in more biologically grounded models of collective decision-making}

\begin{figure*}[htb!]
    \centering
    \includegraphics[width=0.85\linewidth]{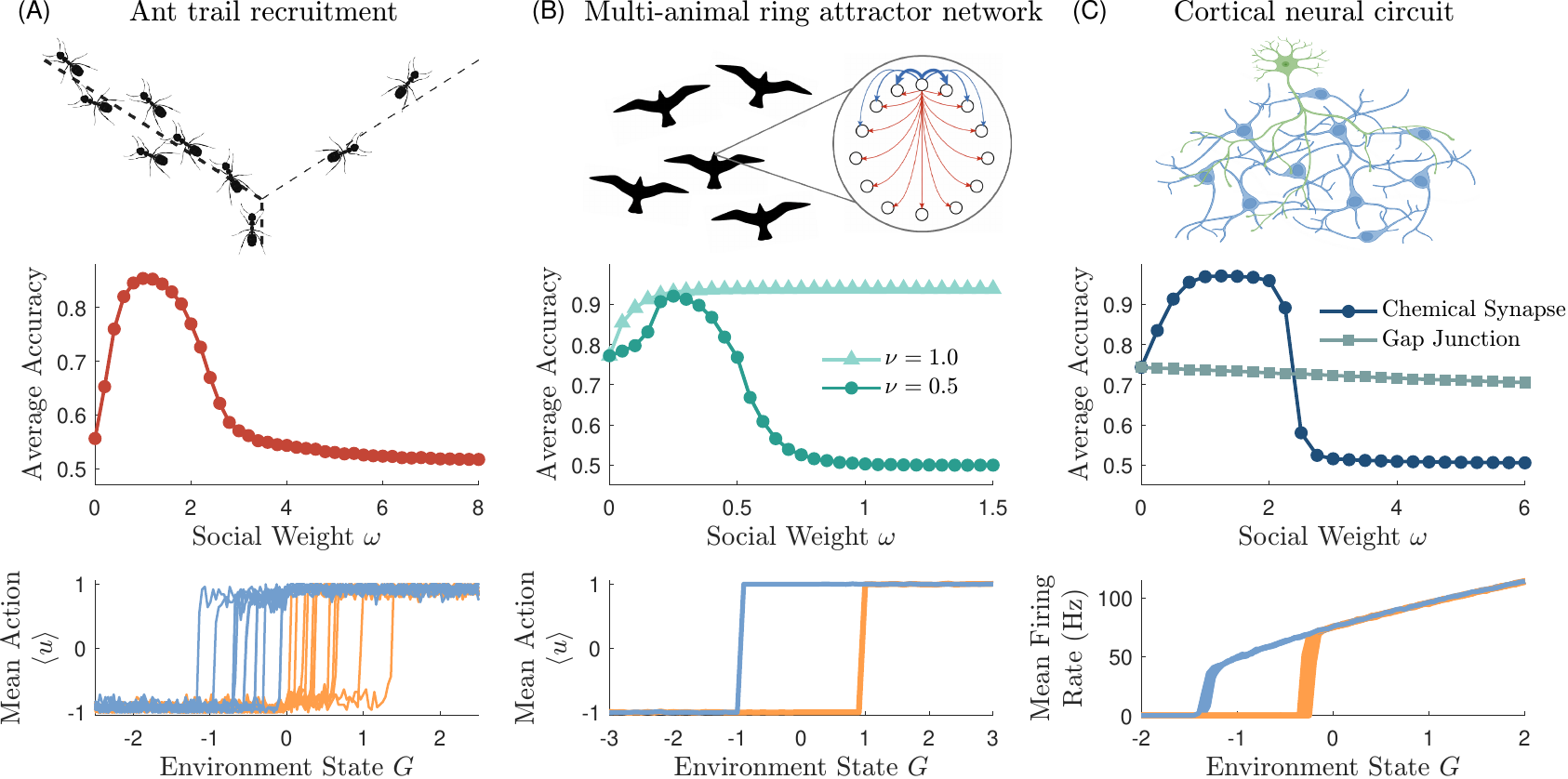}
    \caption{ \textbf{Additional examples of models with greater biological grounding.}
    \textbf{(A)} An ant trail recruitment model.
    \textbf{(B)} A group of animals, each equipped with a ring attractor neural network to aggregate social information, collectively tracks a directional variable.
    \textbf{(C)} A model of a cortical neural circuit collectively tracking a binary environmental state.
    The middle panels show the average decision accuracy as a function of social weight for each model, where a drop in accuracy at high social weight is observed in all three models.
    The bottom panels show hysteresis loops in the collective decision-making dynamics of each model. The ant trail recruitment model shows a higher level of random fluctuation given the stochastic nature of the individual decision-making rule.
    }
    \label{fig:biomodels}
\end{figure*}

The quantized message and limited attention models are intentionally generic, designed to isolate the consequences of two broadly relevant communication constraints. However, collective decision-making in real biological systems involves additional mechanistic details such as pheromone-mediated recruitment, recurrent neural dynamics and spike-based synaptic transmission that complicate these generic models. A natural question is whether the echo chamber dynamics and lock-in arise in more biologically grounded models that include additional realism. To address this, we analyzed three models grounded in specific biological systems: an ant trail recruitment model inspired by pheromone-mediated foraging in ant colonies~\cite{deneubourg1990self,beckers1993modulation,jackson2007modulation,perna2012individual}, a model of directional collective decision-making among animals equipped with ring attractor neural circuits~\cite{sridhar2021geometry,sayin2025behavioral,gorbonos2024geometrical,oscar2023simple}, and a recurrent cortical circuit model~\cite{brunel2000dynamics,wang2002probabilistic,dayan2005theoretical} in which neurons track a binary environmental state and communicate via either spike-mediated chemical synapses or continuous electrical coupling through gap junctions. (Models described in SI Sections \ref{sec:SI_ant}, \ref{sec:SI_ring}, \ref{sec:SI_neural}). In these models, social weight occurs as a key parameter that determines the strength of coupling between group members.

All three models reproduce the same qualitative pattern observed in the generic models (Fig.~\ref{fig:biomodels}): as social weight increases from zero, collective accuracy initially improves, then collapses abruptly beyond a critical threshold. Moreover, all three models exhibit multistability and hysteresis at high social weights, confirming that the echo chamber–like effects identified in our generic framework are not artifacts of simplified communication rules but persist in mechanistically richer settings.

Crucially, each biological model can be understood through the encoding–aggregation decomposition introduced above (Eq.~\ref{eq:general_fg}), revealing a direct mechanistic correspondence with the generic quantized message and limited attention models. In the ant trail recruitment model, agents communicate indirectly through shared pheromone fields. Each agent's choice of foraging trail and the deposited pheromone are naturally quantized messages.
The aggregation of these signals occurs through a nonlinear pheromone response function that maps accumulated pheromone levels to choice probabilities. The nonlinearity in this response function amplifies differences between the two options, creating positive feedback. This self-reinforcing loop creates stable fixed points away from the true environmental state.

In the multi-animal ring attractor model, the nonlinearity resides not in encoding but in aggregation. When the interaction kernel is broad ($\nu = 1$), each agent's ring attractor circuit averages directional inputs linearly, and no bifurcation occurs. This scenario is analogous to the DeGroot-style model. When the kernel is sharp ($\nu < 1$), the circuit implements soft winner-take-all, selectively attending to a subset of inputs while suppressing others. This mechanism provides a biologically grounded instantiation of a nonlinear aggregation function.

In the cortical circuit model, the same recurrent network of neurons exhibits qualitatively different dynamics depending solely on how neurons communicate. Spike-mediated chemical synapses transmit all-or-nothing action potentials as a naturally quantized messages and produce multistability and hysteresis at high coupling strengths. Gap junctions directly share continuous membrane potentials, implementing linear state-sharing, and do not produce these phenomena. This comparison isolates the role of message quantization from other circuit properties. We note that our single-population model deliberately omits the mutual inhibition between competing neural populations that is characteristic of cortical decision-making circuits~\cite{wang2002probabilistic}. In such circuits, a competing population can recruit feedback inhibition to override a locked-in state, facilitating decision reversal. The lock-in we observe is therefore a direct consequence of encoding nonlinearity unmitigated by any competitive reset mechanism, suggesting that mutual inhibition in decision circuits may serve not only to implement winner-take-all selection but also to preserve the network's capacity to revise its collective state when the environment changes.

\section*{Strategies for circumventing echo chambers}

\subsection*{The importance of low in-degree agents}

By including more realistic constraints on how agents send and interpret messages, we found that collectives can experience catastrophic declines in performance when agents over-rely on socially-transmitted messages. But notably, if social weights are tuned to intermediate values, the performance of collective decision-making can be high and comparable to what agents could achieve if they were able to directly pool their direct observations or internal estimates (Fig.~\ref{fig:sw}). We found similar narrow optima in more biologically detailed models of collective decision-making (Fig.~\ref{fig:biomodels}). One interpretation of this result is that biological systems could achieve high performance by precisely tuning social weights for a given collective task. However an alternative is that other mechanisms may allow collectives to achieve high performance without precise tuning of individual behavior. 

\begin{figure*}[htb!]
    \centering
    \includegraphics[width=0.9\linewidth]{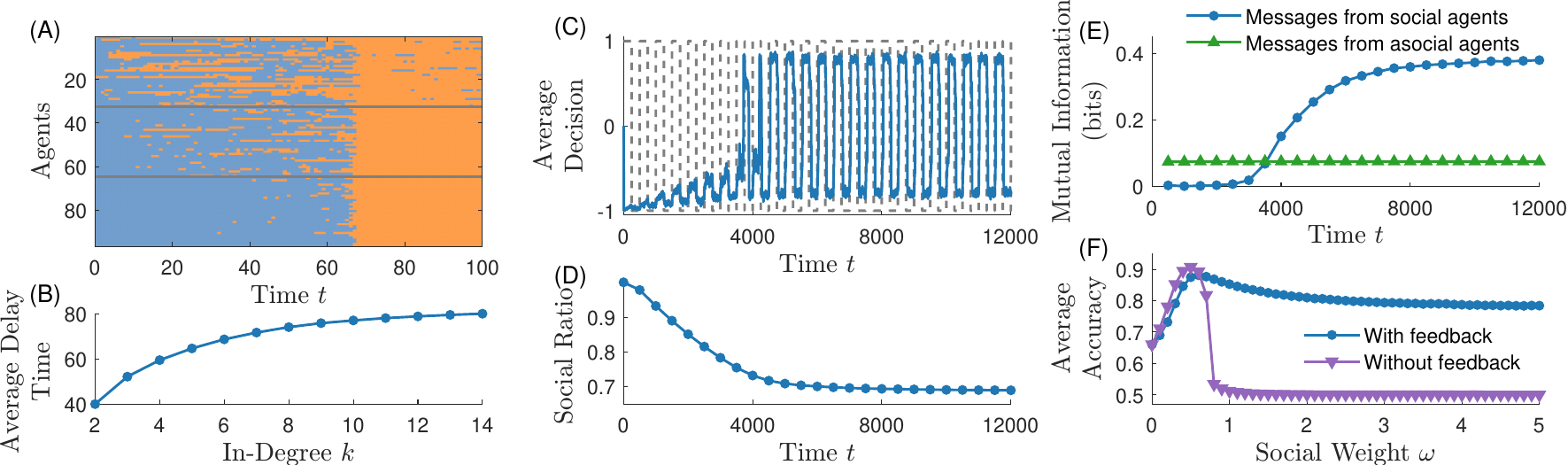}
    \caption{\textbf{Payoff-related feedback can eliminate echo chamber-like states.}
    \textbf{(A)} Response of agents in an ER network with three different levels of in-degree to an environmental change. Agents are sorted in order of increasing in-degree (low to high). The agents shown in the top, middle, and bottom fractions have, respectively, an average in-degree of $\langle k \rangle=7.6, 16.0$, and 26.5. At time $t=0$ (vertical red dashed line), the environment flips from $G_t=-1$ to $G_t=+1$. Agents with a lower in-degree tend to respond earlier.
    \textbf{(B)} Average delay in response versus the in-degree of the agents. This curve comes from repeating the simulation in panel (A) on 500 randomly generated ER networks.
    \textbf{(C)} In a binary message model with $\omega=4$ on a ER network with $N=512$ and $\langle k \rangle=8$, the system would normally enter a locked-in state. Introducing a feedback mechanism that allows agents to switch between social and asocial states (with all agents initially social) leads to a transient period during which the fraction of social agents relaxes to an equilibrium value. This process enables the system to escape the locked-in state and regain sensitivity to environmental flips.
    \textbf{(D)} Time evolution of the fraction of social agents, averaged over 4,000 simulations like the one in panel (C).
    \textbf{(E)} Time evolution of the average mutual information between social messages and the ground truth $G_t$ in the binary message model with feedback. The two curves correspond to messages from social (blue dots) or asocial (green triangles) agents. While the mutual information with messages from asocial agents remains a constant, the mutual information with messages from social agents increases during the transient phase and saturates after the system reaches equilibrium. The presented curves are averaged over 4,000 independent trials.
    \textbf{(F)} Average accuracy as a function of social weight in the binary message model with or without the feedback mechanism. With the feedback, although the peak accuracy is slightly lower, there is no sudden drop caused by hysteresis or locked-in dynamics.}
    \label{fig:network}
\end{figure*}

To identify possible mechanisms, we studied the microscopic dynamics of collective decisions in the model with binary messages as the system approaches a critical value of $\omega$ at which a bifurcation occurs (Fig.~\ref{fig:sw}B). In this regime, changes in the mean action of agents in the collective lag changes in the environmental state.  We find that, when the environment changes, the first agents to change their estimates and corresponding actions to match the new state are agents with low in-degree $k$~
(Fig.~\ref{fig:network} A-B). Let us assume that, before the environment changes, the majority of agents have accurate estimates of the environment's state. Then the environment changes state. For an agent to be among the initial responders, its personal observation must satisfy $|o_{i,t}|>\omega k_i$, as derived from \eqref{eq:gen_model_e}. When $k_i$ is small, this threshold is lower, meaning that low in-degree agents require lower magnitude observations in order to decide to change their estimates and corresponding actions. These low in-degree agents are followed by intermediate degree agents, and finally high degree agents (\ref{fig:network}A,B).

\subsection*{Payoff-sensitive attention} 
In-degree is one measure of the degree to which an agent is influenced by social messages. Given our observation that agents with different in-degrees seem to serve different roles in collective decision-making, we wondered whether individual agents could adopt strategies that dynamically modify in-degree, and whether such a strategy could enable them to benefit from social learning without suffering the risk posed by mis-tuning of social weight. One strategy that has been proposed to mitigate risks of using socially-acquired information to make decisions is for individuals to cease using messages from others when the outcome of decisions is poor, a strategy that has been referred to as ``social learning with payoff-sensitive reinforcement" \cite{aplin2017conformity}. Aplin and colleagues \cite{aplin2017conformity} found that foraging birds are influenced by the foraging decisions of other birds when deciding which sites to explore, but abandon this social influence if they receive poor payoffs from copying others. This strategy belongs to a broader class of context-dependent social learning strategies \cite{Torney2009,aplin2017conformity}.

To incorporate such behavior, we begin with the quantized messages model with binary messages. Each agent is allowed to be in one of the two modes: a social mode or an asocial mode. In the social mode, the agent behaves just like in our standard binary message model. In the asocial mode, the agent sets $\eta_{ij,t}=0, \, \forall \, j,t$ (see \eqref{eq:gen_model_e}), 
meaning it completely ignores all social messages, which is equivalent to having an in-degree of zero. In this asocial mode, agents rely only on their own private noisy observations of the environment.
We implement payoff-sensitive attention through occasional performance feedback. At each time step, after agents have selected their actions $u_i(t)$, each agent independently receives feedback with probability $p_{\mathrm{check}}\ll 1$, which we call the check rate. 
The feedback lasts for one time step and reveals only whether the agent's action matches the true environmental state, i.e., whether $u_i(t)=G_t$. In a real biological setting, such a mechanism could be realized, for example, if individuals periodically evaluate rewards they receive as could occur during behaviors like foraging. If an agent receives feedback indicating that its action is incorrect ($u_i(t)\neq G_t$), it switches its mode (from social to asocial, or vice versa). The agent then remains in the new mode until it next receives feedback indicating an incorrect action. 
The check rate $p_{\mathrm{check}}$ is assumed to be small (e.g., $p_{\mathrm{check}} = 2\times10^{-4}$ in Fig.~\ref{fig:network}), so that such feedback events are rare. 

We find that this adjustment to individual behavior qualitatively changes dynamics (Fig.~\ref{fig:network}C-F). Under a social weight that would result in strong hysteresis and degraded performance under the standard quantized message model ($\omega =4$), we find that agents initially perform poorly at tracking the environment (Fig.~\ref{fig:network}C), but over time, as some agents receive feedback that they have selected the wrong action, the fraction of the population with non-zero in-degree decreases (Fig.~\ref{fig:network}D), and the population begins to accurately track the environmental state (Fig.~\ref{fig:network}C, $t > 4250$). The mutual information between the true environmental state, $G_t$, and the typical messages shared by agents within the network increases as a subset of agents begin to ignore their neighbors (Fig.~\ref{fig:network}E). Somewhat counterintuitively, the agents that benefit most from this are the agents that continue to attend to neighbors (i.e,. those with in-degree $> 0$; Fig.~\ref{fig:network}E, blue curve), which exhibit states and therefore messages that contain more information about the true state of the environment than do agents that ignore messages. 
This echoes a finding by Ref.~\cite{madhushani2019heterogeneous} in the context of multi-agent bandit problems, where agents perform best when they have high sociability and their neighbors have low sociability.
This phenomenon allows collectives to maintain high accuracy across a broad range of social weights (Fig.~\ref{fig:network}F), eliminating the need to precisely tune individual behavior.

\section*{Conclusions}

In this work, we have framed the problem of collective decision-making in a dynamic environment in a general way that encompasses both widely-studied mathematical models like the DeGroot-style and wisdom of the crowds models, as well as models that capture two pervasive features of collective decision-making in real biological systems: message quantization, which occurs when agents transform their internal states or beliefs into discrete actions that are then observed by others \cite{fahimipour2023wild,giraldeau2002potential}, and limited attention, which results in agents basing their actions, and therefore the messages they convey to others, on only a subset of the messages they receive \cite{greenfield2021rhythm,lemasson2018motion,lemasson2009collective,van2010selective}. 

Contrary to the prediction that discretizing message space to a small number of possible messages or limiting communication to a small number of possible channels may simply degrade communication~\cite{msechu2008decentralized,nedic2009distributed,li2010distributed}, we find that these constraints cause a qualitative change in the dynamics of collective decision making. This qualitative change can be characterized mathematically by the appearance of bifurcations that lead to multistability and hysteresis. When individuals within a collective over-rely on socially-transmitted messages, the system can cross this bifurcation opening the possibility of becoming locked in a state in which agents send social messages that are at odds with the true state of the environment, even as each agent makes observations that indicate the environmental state has changed. This sensitive dependence on the weights agents apply to messages from their neighbors means that, for a given collective task, agents within the network must tune their social weights within a fairly narrow range to balance robustness to noise against the risk of insensitivity (Fig.~S\ref{fig:addEnv}). 

Prior work showed that quantized communication can induce nonconsensus equilibria on general graphs~\cite{ceragioli2018consensus}. Here we identify a different failure mode: system-wide breakdown of environment tracking that can persist even on fully connected networks. Communication constraints may therefore give rise to at least two classes of collective failure with distinct mechanisms (see SI Section \ref{sec:SI_addnets}).

Past theoretical work on models with binary messages has shown that natural selection on individual performance can push social weights beyond their optimal values (peaks in Fig.~\ref{fig:sw}A) and into a regime where the collective becomes insensitive to changes in the environment  \cite{torney2015social}. Our work suggests that the existence of a unimodal relationship between mean accuracy of the group and social weight with a steep decline in accuracy beyond the optimum is not a feature of this particular model, but rather is a very general feature of collectives in which messages are quantized versions of agents' internal states or agents have limited attention. Importantly, these features also emerge in models developed to describe collective decision-making in specific biological systems (Fig. \ref{fig:biomodels}).

More generally, our analysis reveals that the breakdown of collective tracking can arise from two distinct but complementary mechanisms. At the dynamical-systems level, collective lock-in may emerge through changes in the number of stable fixed points via bifurcations, or through constraints on the reachability of those fixed points in dynamic environments. At the level of information processing, these mechanisms correspond to nonlinearities introduced either in message encoding, when agents transform internal estimates into discrete or coarse-grained signals, or in message aggregation, when agents selectively integrate only a subset of incoming messages. These information-processing nonlinearities can in turn produce collective distortion, such that socially transmitted messages no longer faithfully track the environmental state. This perspective suggests that these failure modes are not model-specific, but a general consequence of constraints on how information is transmitted and combined within a group. As a result, diverse biological and social systems, despite differences in their underlying mechanisms, may exhibit similar breakdowns in collective decision-making.


Importantly, features of both messaging and network topology can shape the severity of echo chamber dynamics. We find that low-degree agents are particularly important in triggering responses to changes in the state of the environment, and collectives with communication network topologies that have a large fraction of such agents, or lack cycles altogether, are less prone to becoming unresponsive to changes in the environment (Fig.~S\ref{fig:SI_heterNets}). In the case of quantized messages, the richness of the message set (i.e., the number of distinct messages an agent can receive) governs the severity of hysteresis, such that collective distortion is most severe for binary message sets and becomes less severe as the message set becomes richer. This observation suggests that, even if messages are produced by discrete actions, agents that observe these actions can benefit by seeking to discern variation in actions that may convey additional information about neighbors' internal states. There is evidence that apparently discrete actions such as fleeing from a putative threat contain significant variation in vigor that is correlated with the severity of the threat \cite{evans2018synaptic,evans2019cognitive}. Moreover, when nearby individuals observe an animal fleeing, the strength of the stimulus produced by the fleeing individual is positively correlated with the vigor of its response \cite{fahimipour2023wild}. These empirical findings suggest that it may be possible for organisms to extract additional information from apparently quantized messages. Incorporating such variation would effectively enrich the message set and reduce the risk of collective distortion.

Dynamic adjustment of social influence offers another route to robustness. The payoff-dependent attention strategy we analyzed enables agents to break echo chambers by temporarily ignoring social messages following negative feedback, thereby destroying the multistability that underlies lock-in (Fig.~\ref{fig:network}F). This finding aligns with empirical evidence that animals modulate social influence in response to experienced payoffs \cite{aplin2017conformity} and with theoretical proposals for dynamic gain modulation in collective systems \cite{leonard2024fast}.

In this work, we show that constraints on how organisms perceive and process information from their social partners can dramatically affect the quality of collective decision-making by qualitatively changing the dynamics of information flow within a collective. The theoretical framework developed here provides a basis from which to understand phenomena related to collective decision-making across many areas of biology.

\section*{Acknowledgment}
This manuscript benefitted from conversations with BT Martin, S Strogatz, J Guckenheimer, and members of the Hein and Leonard laboratories. A.M.H. acknowledges the National Science Foundation (IOS-2338596, and EF-2222478). L.-W.K. acknowledges funding from the Eric and Wendy Schmidt AI in Science Postdoctoral Fellowship, a Schmidt Futures program.

\newpage
\onecolumngrid
\appendix

\section{Mean-field theories and bifurcation analysis}\label{sec:SI_mf}

As discussed in the main text, we develop mean-field theories for collective decision-making models studied in this work. This section provides the detailed derivations and additional mean-field results for these models. For the observation-sharing model, the dynamics have a simple behavior so a separate mean-field closure is not required and we thus focus on DeGroot-style, quantized message, and limited attention models only.

The mean-field theories for the three models rely on the same set of assumptions stated in the main text. At each time $t$, we approximate the population of internal estimates $\{e_{i,t}\}$ as i.i.d.\ Gaussian random variables $e_{i,t} \sim \mathcal{N}(\mu_t,\sigma_t^2)$, where $\mu_t=\langle e_t\rangle$ and $\sigma_t$ denote the mean and standard deviation, respectively. We further assume that observational noise terms are i.i.d.\ across agents and time. Under these assumptions, the central task of the mean-field analysis is to derive an iterative map $\{\mu_{t-1},\sigma_{t-1}\}\to\{\mu_t,\sigma_t\}$, which reduces the high-dimensional agent-based stochastic dynamics to a low-dimensional deterministic map.

\subsection{General explicit form of the mean-field theory for the quantized message model}

In the main text, we briefly discussed the mean-field theory for the quantized message model with binary messages. Here we present the general formulation for an arbitrary finite set of quantized messages 
\begin{align}  
Q = \{q_j \in \mathbb{R} \mid j = 1,2,\dots,N_Q\}.
\end{align}
Without loss of generality, assume
\begin{align}  
q_1 < q_2 < \cdots < q_{N_Q},
\end{align}
and denote the quantization thresholds by
\begin{align} 
-\infty = b_0 < b_1 < \cdots < b_{N_Q} = \infty,
\end{align}
where the quantization function $f_Q(x)$ maps $x \in (b_{j-1}, b_j]$ to $q_j$.

Under the mean-field assumption that the pre-quantized internal estimates are Gaussian distributed as $\mathcal{N}(\mu_{t-1}, \sigma_{t-1}^2)$, the mean and variance of the social messages are given by
\begin{align}
M_t^{(1)} 
&= \sum_{j=1}^{N_Q} q_j 
\left[
\Phi\!\left(\frac{b_j - \mu_{t-1}}{\sigma_{t-1}}\right)
-
\Phi\!\left(\frac{b_{j-1} - \mu_{t-1}}{\sigma_{t-1}}\right)
\right], \\
M_t^{(2)} 
&= \sum_{j=1}^{N_Q} q_j^2 
\left[
\Phi\!\left(\frac{b_j - \mu_{t-1}}{\sigma_{t-1}}\right)
-
\Phi\!\left(\frac{b_{j-1} - \mu_{t-1}}{\sigma_{t-1}}\right)
\right]
- \big(M_t^{(1)}\big)^2,
\end{align}
where $\Phi(\cdot)$ denotes the standard normal cumulative distribution function.
The mean and variance of the estimates then evolve according to
\begin{align}
\mu_t &= (1-\omega_s) G_t + \omega_s M_{t-1}^{(1)}, \\
\sigma_t^2 &= (1-\omega_s)^2 \sigma_n^2 
+ \omega_s^2 \frac{M_{t-1}^{(2)}}{k},
\end{align}
where $\sigma_n = \sigma_{OU}/\sqrt{2\omega_g}$ is the standard deviation of the observational noise, 
$k$ is the number of incoming social messages, and $\omega_s = 1 - 1/(k\omega + 1)$ is the effective total social weight. Together, these equations define the complete iterative mean-field map for generic quantized-message models.

\subsection{Bifurcation analysis of the mean-field theory for the quantized message model}
We now show that the bifurcations observed in the quantized message model are governed by (unfolded) pitchfork bifurcations.

\begin{figure*}
    \centering
    \includegraphics[width=0.9\linewidth]{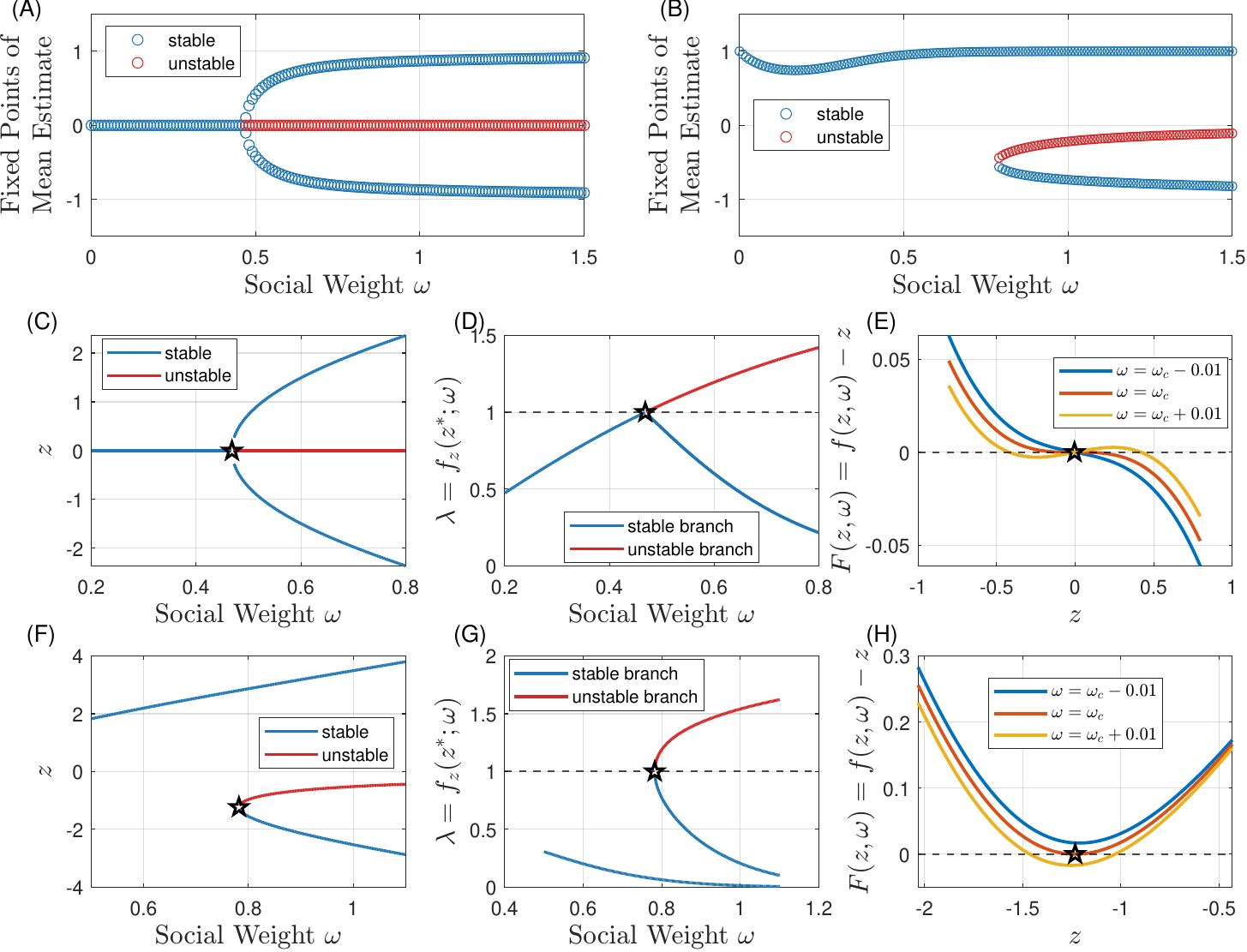}
    \caption{Analysis of the bifurcation in the binary message model.
    \textbf{(A)} Bifurcation diagram of the mean estimate in a constant neutral environment $G_t=0$.
    \textbf{(B)} Bifurcation diagram of the mean estimate in a constant environment $G_t=+1$. The diagram for $G_t=-1$ has the same structure but is inverted due to the symmetry of the system.
    \textbf{(C, D, E)} Bifurcation analysis of the reduced one-dimensional map for $z$ in a neutral constant environment $G_t=0$. 
    (C) Bifurcation diagram. (D) Eigenvalues of the Jacobian for each branch, indicating the stability of the fixed points. (E) Zero crossings of $F(z)=f(z)-z$ near the bifurcation point $\omega_c$.
    \textbf{(F, G, H)} Bifurcation analysis of the reduced one-dimensional map for $z$ in a constant environment $G_t=+1$. 
    (F) Bifurcation diagram. (G) Eigenvalues of the Jacobian for each branch, indicating the stability of the fixed points. 
    (H) Zero crossings of $F(z)=f(z)-z$ near the bifurcation point $\omega_c$. $k=6.82$ and $\sigma_n=2.24$ for all panels. }
    \label{fig:BifurJaco_Qb}
\end{figure*}

We first use an example of the binary message scenario $Q=\{-1,+1\}$. As discussed in the main text, the iterative map $\{\mu_{t-1},\sigma_{t-1}\}\to\{\mu_t,\sigma_t\}$ is as follows:
\begin{align}
\mu_t &= (1-\omega_s) G_t + \omega_s [2 \Phi(\frac{\mu_{t-1}}{\sigma_{t-1}}) -1],\\
\sigma_t^2 &= (1-\omega_s)^2 \sigma_n^2 
+ \frac{\omega_s^2}{k} \{1- [2 \Phi(\frac{\mu_{t-1}}{\sigma_{t-1}}) -1]^2\}.
\end{align}
Note that this map can be reduced to a one-dimensional map. Take $z_t=\mu_t/\sigma_t$, we have
\begin{align}
\mu_t &= (1-\omega_s) G_t + \omega_s [2 \Phi(z_{t-1}) -1],\\
\sigma_t^2 &= (1-\omega_s)^2 \sigma_n^2 
+ \frac{\omega_s^2}{k} \{1- [2 \Phi(z_{t-1}) -1]^2\}.
\end{align}
The iterative map for $z_t$ is
\begin{align}
    z_t &= \frac{\mu_t}{\sigma_t}  \\
        &= \frac{(1-\omega_s) G_t + \omega_s [2 \Phi(z_{t-1}) -1]}
        {\sqrt{(1-\omega_s)^2 \sigma_n^2 
+ \frac{\omega_s^2}{k} \{1- [2 \Phi(z_{t-1}) -1]^2\}}} \label{eq:zmap}.
\end{align}
Denote this iterative map by $z_t = f(z_{t-1})$, and define $F(z) = f(z) - z$. Fixed points $z^*$ satisfy $F(z^*) = 0$. Let $\omega_c$ denote the critical social weight at which a bifurcation occurs. In Fig.~\ref{fig:BifurJaco_Qb}, we show the stable and unstable branches of this map system, under environment states $G_t=0$ or $G_t=1$, with corresponding eigenvalues. We show that, when the environment is neutral $G_t=0$, the system undergoes a supercritical pitchfork bifurcation, where one stable branch becomes one unstable branch and two symmetric stable branches at the bifurcation point $\omega_c$. When the environment is $G_t=1$, the symmetry is broken, and the bifurcation becomes an unfolded pitchfork bifurcation, which is also a type of fold bifurcation.


\begin{figure*}
    \centering
    \includegraphics[width=0.9\linewidth]{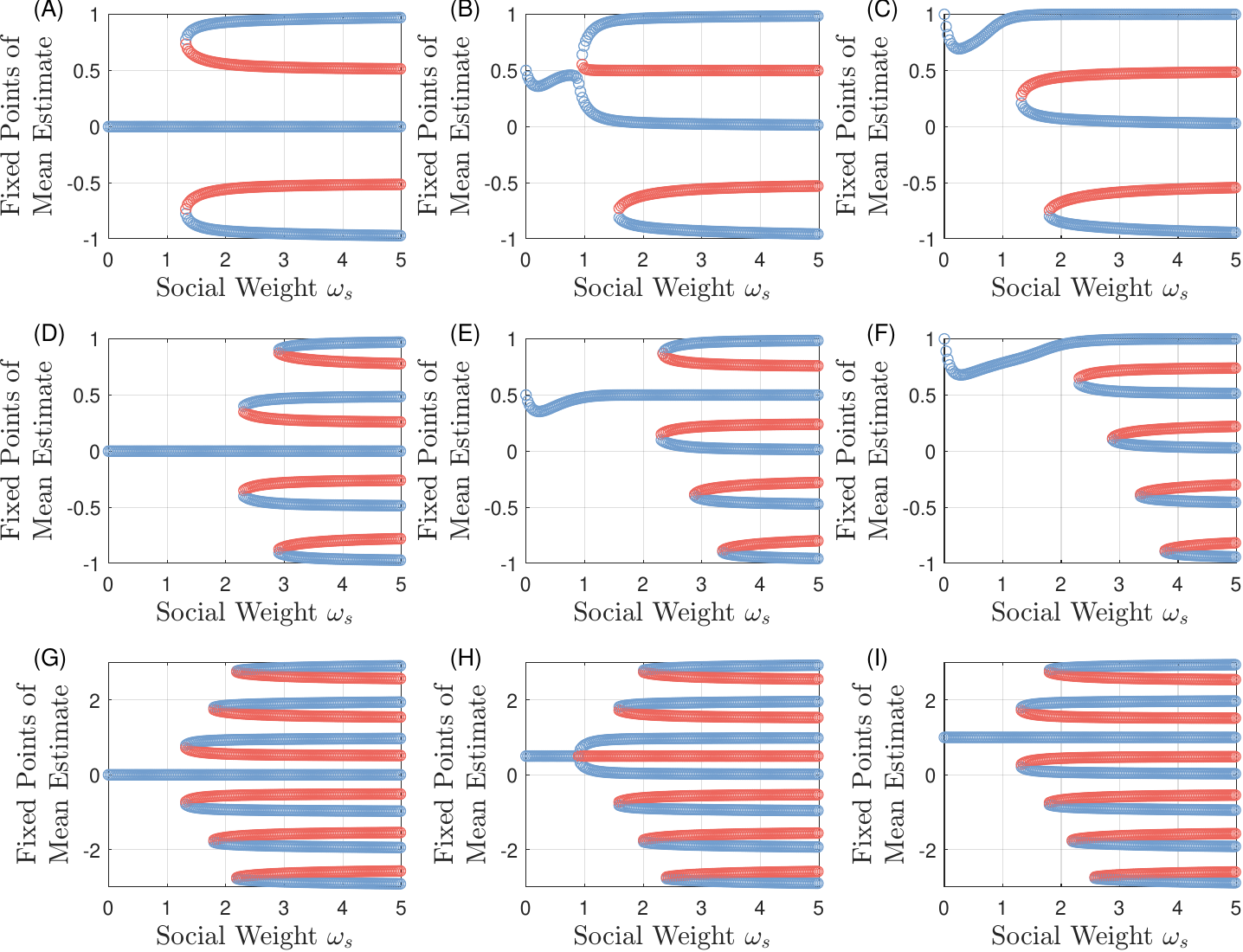}
    \caption{Bifurcation diagrams of different quantization sets $Q$ under different constant environments $G$. Different rows correspond to different $Q$ and different columns correspond to different $G$. For panels in the top row (A, B, C), $Q=\{-1,0,1\}$. For panels in the middle row (D, E, F), $Q=\{-1,-0.5,0,+0.5,+1\}$. For panels in the lower row (G, H, I), $Q=\mathbb{Z}$. For panels in the left column (A, D, G), $G=0$. For panels in the middle column (B, E, H), $G=0.5$. For panels in the right column (C, F, I), $G=1.0$. We use $1/\langle 1/k\rangle =6.82$ and $\sigma_n=2.24$ for all panels. Here, we focus on the bifurcation structures in constant environments. Some portions of the stable solutions may not be reachable in a switching environment.}
    \label{fig:Bifur_Q}
\end{figure*}

\begin{figure*}
    \centering
    \includegraphics[width=0.6\linewidth]{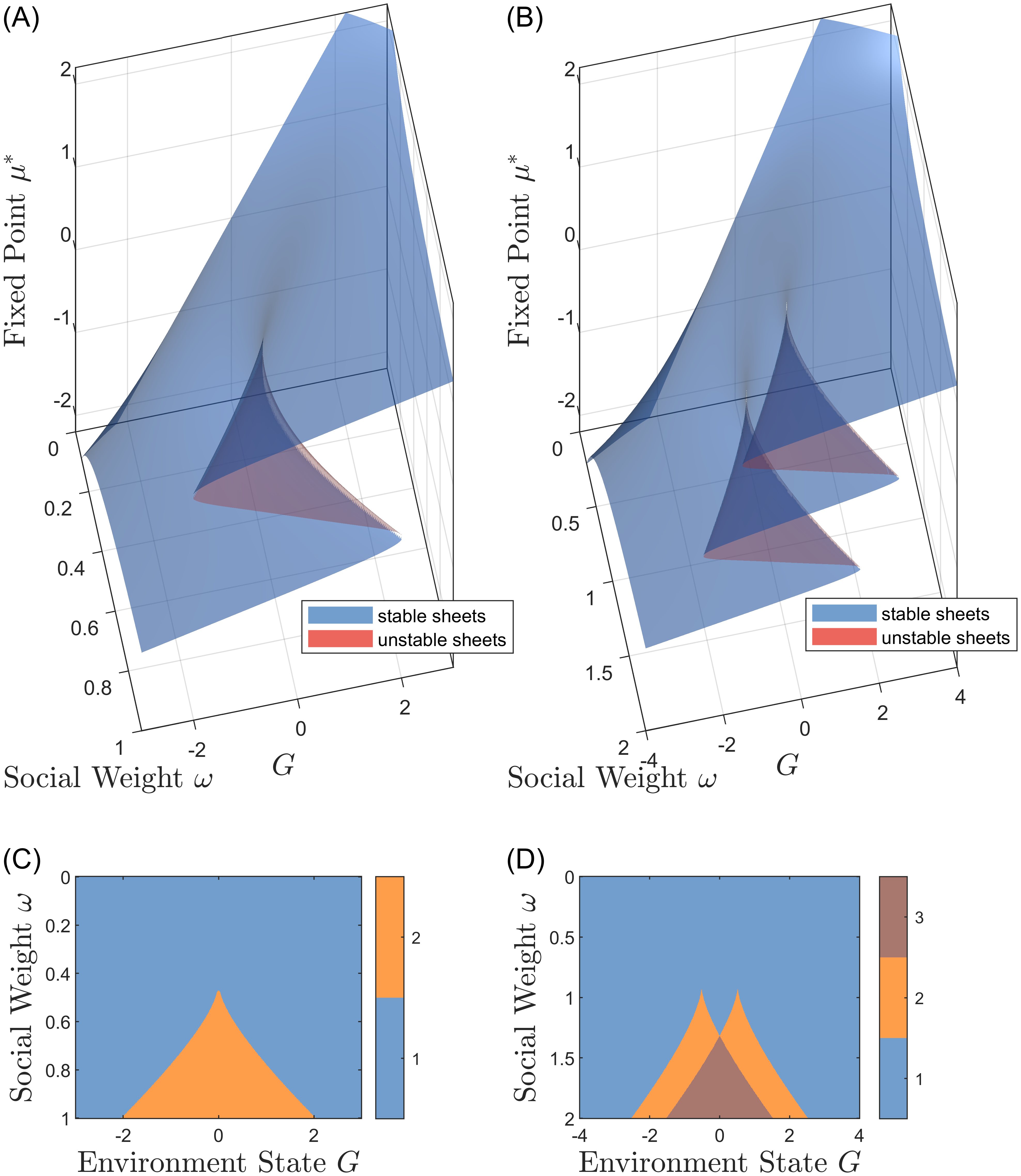}
    \caption{Cusp bifurcation in quantized message model with $Q=\{-1,+1\}$ (A, C) and  $Q=\{-1,0,+1\}$ (B, D). The upper panels (A,B) show the surface of fixed points in the two-parameter $(\omega, G)$ space. The lower panels (C, D) show the number of stable fixed points at different $\omega$ and $G$.}
    \label{fig:SI_Cusp}
\end{figure*}

In Fig.~\ref{fig:Bifur_Q}, we show additional examples of bifurcation diagrams for different quantization sets $Q$. A general trend can be observed as the size of the quantization set $Q$ increases. When $Q$ contains more discrete values, additional stable fixed points appear at large social weight $\omega$. Since the system has only a single stable fixed point at $\omega=0$, these additional stable states must be created through bifurcations as $\omega$ increases. In particular, they arise through fold bifurcations or supercritical pitchfork bifurcations that generate new pairs of stable and unstable fixed points.
The phenomena observed here are consistent with a cusp bifurcation~\cite{zeeman1976catastrophe,kuznetsov1998elements}. Treating $\omega$ and $G$ as bifurcation parameters, we can visualize the surface of fixed points as shown in Fig.~\ref{fig:SI_Cusp}(A,B). This surface exhibits folded structures, and increasing the number of elements in the quantization set $Q$ leads to an increase in the number of folds. Figure~\ref{fig:SI_Cusp}(C,D) show the number of stable fixed points across parameter space. By traversing the two-dimensional $(\omega, G)$ parameter plane at different fixed values of $G$, we obtain different one-dimensional bifurcation diagrams, from the ones illustrated in Fig.~\ref{fig:Bifur_Q}.

All the diagrams shown in this section are obtained under a constant environment and therefore do not include the effect of environmental switching. Consequently, they differ from the bifurcation structures in Fig.~3 of the main text, where the environment switches over time. When environmental switching is taken into account, some stable branches become dynamically unreachable. As discussed in the main text, the collective decision-making dynamics can be viewed as a switched dynamical system, where each constant environment state corresponds to a subsystem. The bifurcation structures analyzed in this section, therefore, correspond to those of the individual subsystems.

\subsection{Mean-field theory for the DeGroot-style model}
We can derive the mean-field theory for the DeGroot-style model by replacing the quantization function $f_Q(\_)$ by an identity function. Thus, the iterative map is as follows.
\begin{align}
    \mu_t &= (1-\omega_s) G_t + \omega_s \mu_{t-1},\\
    \sigma_t^2 &= (1-\omega_s)^2 \sigma_n^2 + \omega_s^2 \frac{\sigma_{t-1}^2}{k}.
\end{align}
Note that the two dimensions are independent. $\mu_t$ does not depend on $\sigma_{t-1}$, and $\sigma_t$ does not depend on $\mu_{t-1}$. There is only one solution taking $\mu_{t-1}=\mu_t$ and $\sigma_{t-1}=\sigma_t$, as follows, which is a stable fixed point.
\begin{align}
    \mu_t &= G_t,\\
    \sigma_t^2&= \frac{(1-\omega_s)^2}{1-\omega_s^2/k} \sigma_n^2.
\end{align}
This stable fixed point is globally attracting for any initial condition, and there is no bifurcation in the system as one changes the social weight. Thus, there is no multistability or hysteresis, and there are no phenomena of locked-in or long delays. 

\subsection{Mean-field theory for the limited attention model}

We develop a closed-form mean-field theory for the limited attention model. For each agent, while the social messages from $k_i$ neighbors are available, it only takes in the $L$ social messages with the largest absolute magnitude $|m|$ and aggregates these $L$ social messages with the personal observation. For simplicity, we assume the in-degree across the network is uniformly $k$, and focus on the scenarios where $k>L$.

Under a mean-field assumption that the $k$ incoming social messages to an agent are i.i.d.\ draws from this distribution, our top-$L$ selection by magnitude can be approximated by \emph{two-sided truncation}: a social message is selected if and only if $|m|>q$, where the threshold $q\ge 0$ is chosen so that the selected fraction matches the expected Top-$L$ rate
\begin{align}
\frac{L}{k} = \Pr(|m|>q).
\end{align}
For a Gaussian $m=e\sim\mathcal{N}(\mu,\sigma^2)$ this implies
\begin{align}
\frac{L}{k} = 1-\Bigg[\Phi\!\Big(\frac{q-\mu}{\sigma}\Big)-\Phi\!\Big(\frac{-q-\mu}{\sigma}\Big)\Bigg],
\label{eq:topk_threshold}
\end{align}
where $\Phi(\cdot)$ is the standard normal CDF. Using Eq.~\ref{eq:topk_threshold}, we can numerically solve the value of $q$, as the values of all the other variables and coefficients are known. We also define the standardized endpoints
\begin{align}
a=\frac{-q-\mu}{\sigma},\qquad b=\frac{q-\mu}{\sigma},
\end{align}
and use $\phi(\cdot)$ to denote the standard normal PDF.

Let $\mu_{\mathrm{sel}}$ and $\sigma^2_{\mathrm{sel}}$ denote the conditional mean and variance of selected top-$L$ messages. The conditional mean is the standard truncated-Gaussian tail mean:
\begin{align}
\mu_{\mathrm{sel}}
= \mu + \sigma\,\frac{\phi(b)-\phi(a)}{L/k}.
\label{eq:tail_mean}
\end{align}
To obtain the conditional variance, we compute the tail second moment via the law of total expectation. We first try to calculate $\mathbb{E}\!\left[m^2\,\big|\,|m|\le q\right]$. 
Suppose there is a random variable $x \sim N(0,1)$ that is a standard Gaussian random variable. We have
\begin{align}
    \mathbb{E}\left[x\,\big|\,a\leq x\ \leq b \right] &= \frac{\phi(a)-\phi(b)}{1-L/k},\\
    \mathbb{E}\left[x^2\,\big|\,a\leq x\ \leq b \right] &= 1+ \frac{a\phi(a)-b\phi(b)}{1-L/k},\\
    m &= \mu + \sigma x.
\end{align}
The conditional second moment on the interval $|m|\le q$ is thus
\begin{align}
\mathbb{E}\!\left[m^2\,\big|\,|m|\le q\right]
=\mu^2
+2\mu\sigma\,\frac{\phi(a)-\phi(b)}{1-L/k}
+\sigma^2\left[1+\frac{a\phi(a)-b\phi(b)}{1-L/k}\right].
\label{eq:in_second_moment}
\end{align}
Since $\mathbb{E}[m^2]=\mu^2+\sigma^2$, the tail second moment is
\begin{align}
\mathbb{E}\!\left[m^2\,\big|\,|m|>q\right]
=\frac{(\mu^2+\sigma^2)-(1-L/k)\,
\mathbb{E}\!\left[m^2\,\big|\,|m|\le q\right]}{L/k}.
\label{eq:tail_second_moment}
\end{align}
Finally,
\begin{align}
\sigma^2_{\mathrm{sel}}
=\mathbb{E}\!\left[m^2\,\big|\,|m|>q\right]-\mu_{\mathrm{sel}}^2.
\label{eq:tail_var}
\end{align}
Now we have the expressions of $\mu_{\mathrm{sel}}$ and $\sigma_{\mathrm{sel}}$ as functions of $\mu$ and $\sigma$. 

As discussed in the main text, we model the personal observation as
\begin{align}
o_t = G_t + \zeta_t,\qquad \zeta_t\sim\mathcal{N}(0,\sigma_n^2).
\end{align}
Given the selected-message statistics $(\mu_{\mathrm{sel}},\sigma^2_{\mathrm{sel}})$ computed from $(\mu_t,\sigma_t^2)$ via \eqref{eq:topk_threshold}--\eqref{eq:tail_var}, the agent's updated continuous estimate is the weighted average
\begin{align}
e_{t+1}=\frac{o_t + L \,\omega \,\bar m_{\mathrm{sel},t}}{1 + L \,\omega},
\end{align}
where $\bar m_{\mathrm{sel},t}$ is the average of the $L$ selected social messages at time $t$.
Under the i.i.d.\ mean-field assumption, we have
\begin{align}
\mathbb{E}[\bar m_{\mathrm{sel},t}] = \mu_{\mathrm{sel}},
\qquad
\mathrm{Var}(\bar m_{\mathrm{sel},t}) = \frac{\sigma^2_{\mathrm{sel}}}{L}.
\end{align}
This yields the closed mean-field map
\begin{align}
\mu_{t+1}
&= \frac{G_t+L\,\omega\,\mu_{\mathrm{sel}}(\mu_t,\sigma_t)}{1+L\,\omega},
\label{eq:mf_topl_mu_update}\\[4pt]
\sigma^2_{t+1}
&= \frac{\sigma_n^2+L\,\omega^2\,\sigma^2_{\mathrm{sel}}(\mu_t,\sigma_t)}{(1+L\,\omega)^2}.
\label{eq:mf_topl_sigma_update}
\end{align}
\eqref{eq:mf_topl_mu_update}--\eqref{eq:mf_topl_sigma_update} define a two-dimensional deterministic iteration for the population-level mean and variance of messages/estimates. Steady states correspond to fixed points $(\mu^\ast,\sigma^{\ast 2})$ under constant $G_t\equiv G$, and time-dependent tracking under a switching environment $G_t$ can be studied by iterating the map with the corresponding input sequence. In SI Sec.~\ref{sec:SI_consensus}, we provide further analysis of the stable fixed points in the limited attention model, especially in the limit of large $\omega$.

\subsection{Bifurcation analysis of the mean-field theory for the limited attention model}
\begin{figure*}
    \centering
    \includegraphics[width=0.75\linewidth]{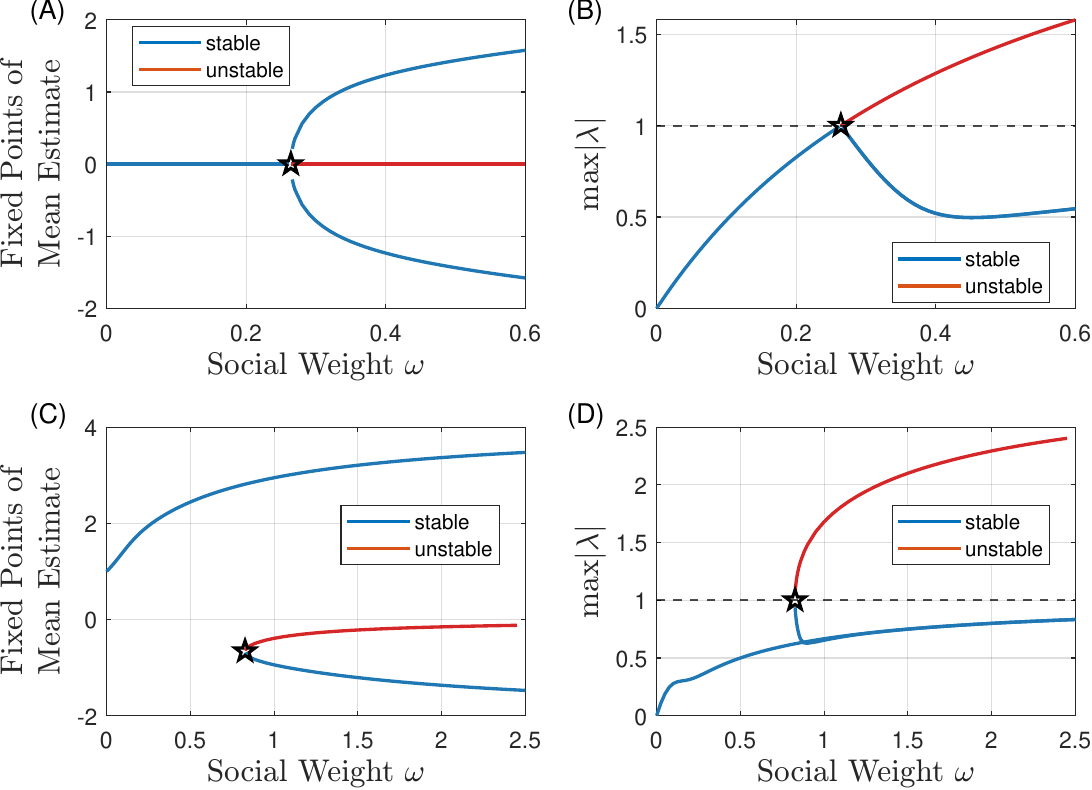}
    \caption{Analysis of the bifurcation in the limited attention model with $L=2$ and $\langle k\rangle=8$.
    \textbf{(A)} Bifurcation diagram of the mean estimate in a constant neutral environment $G_t=0$.
    \textbf{(B)} Eigenvalues of the Jacobian for each branch at $G_t=0$, indicating the stability of the fixed points.
    \textbf{(C)} Bifurcation diagram of the mean estimate in a constant environment $G_t=+1$. The diagram for $G_t=-1$ has the same structure but is inverted due to the symmetry of the system.
    \textbf{(D)} Eigenvalues of the Jacobian for each branch at $G_t=+1$, indicating the stability of the fixed points. In all panels, the unstable branches correspond to saddle nodes. $\sigma_n=2.24$. }
    \label{fig:BifurJaco_TopK}
\end{figure*}

In Fig.~\ref{fig:BifurJaco_TopK}, we show the stable and unstable branches of the limited attention model with $L=2$ and $\langle k\rangle=8$. We also show the corresponding eigenvalues of the Jacobian matrix at the fixed points with the largest absolute values. It appears that when the environment is neutral, there is a supercritical pitchfork bifurcation. When the environment is biased, the bifurcation becomes a fold bifurcation, which, in this case, can also be called an unfolded pitchfork bifurcation.

\section{Relating the original Torney model to the quantized message model.} \label{sec:SI_Torney}

The quantized-message model studied in our work is closely related to, but not identical with, the model introduced by Torney and colleagues~\cite{torney2015social}. Here we describe the main modifications we made in order to place the model within our common framework and facilitate comparison with the other models considered in this paper.

In Ref.~\cite{torney2015social}, the authors focused on the binary-action case and did not consider extensions to richer action sets. Agents observe only the binary actions of their neighbors and use a Bayesian rule to combine this social information with their own noisy personal observations. Importantly, however, agents do not have access to the confidence associated with a neighbor's message, i.e., the probability that the neighbor's action agrees with the true environmental state. The decision rule for an agent's binary action in Ref.~\cite{torney2015social} is
\begin{align}
    u_{i,t} = \text{sign}\!\left(\exp\!\left(\frac{4\omega_g}{\sigma^2} g_i\right)\left(\frac{\omega_s}{1-\omega_s}\right)^{N_+ - N_-} - 1\right).
\end{align}
Here, \(g_i\) denotes the agent's personal observation, \(N_+\) and \(N_-\) are the numbers of neighbors currently taking actions \(+1\) and \(-1\), respectively, and \(\omega_s\) is the agent's assumed probability that a neighbor is correct. In the original paper, \(\omega_s\) is therefore not the true reliability of neighbors, but rather a subjective social-confidence parameter that determines how strongly social observations are weighted in the Bayesian update. As emphasized by Torney et al., agents also cannot know whether different neighbors provide independent evidence or whether some are simply repeating information acquired from others.

Our quantized-message formulation differs from the original Torney model in several ways. First, rather than working directly with the exact Bayesian expression above, we rewrite the update rule in a form that fits the common weighted-combination framework used throughout this paper. This allows us to compare models that differ in how messages are encoded while preserving the same basic architecture: agents combine a noisy personal observation with socially received messages and then emit a message to others. Second, we interpret observed neighbor actions as quantized messages, which makes it straightforward to extend the model beyond the binary case considered by Torney et al. In this sense, the original Torney model can be viewed as a special case of our quantized-message framework with a binary message set.

A further difference is that Torney et al. derived their decision rule under a specific prior assumption about the environmental dynamics, in particular triangle-wave and square-wave environments. As a result, the Bayesian decision boundary depends explicitly on the assumed temporal statistics of the environment. In our framework, by contrast, we do not attempt to preserve exact Bayesian optimality under a particular prior. Instead, we use a simplified message-passing formulation that makes it possible to compare continuous, quantized, and other biologically motivated messaging strategies within a unified dynamical framework.

Despite these differences, the conceptual connection between the two models is close. In both cases, agents observe only neighbors' discrete actions rather than their internal confidence, creating the possibility that socially propagated information becomes over-weighted. This feature is central to the original Torney model, in which increasing social confidence initially improves performance but eventually leads to delayed responses and collective unresponsiveness. Our quantized-message model preserves this same core mechanism while generalizing it and embedding it in a broader comparative framework.

\section{Two routes of collective lock-in: encoding versus aggregation nonlinearity, and the singular $\omega \to \infty$ limit of the limited attention model} \label{sec:SI_consensus}

In the main text, we decompose the collective dynamics in the large-social-weight limit into an encoding function $f$ mapping estimates to outgoing messages and an aggregation function $h$ combining incoming messages (Eq.~\ref{eq:general_fg}). We assume that both $f$ and $h$ are monotonic and odd-symmetric (reflecting the symmetry between the two binary environmental states), and that $h$ is permutation-invariant over its inputs. Below, we analyze how nonzero consensus fixed points arise in the two models studied in this paper.

\subsection{Quantized message model: nonlinearity in encoding}

In the quantized message model, $h$ is linear averaging and $f = f_Q$ is the quantization function. In the limit $\omega \to \infty$, where personal observations vanish, the dynamics on a homogeneous network reduce to
\begin{align}
    e_{i,t} = \frac{1}{k} \sum_{j} A_{ij}\, f_Q(e_{j,t-1}). \label{eq:consensus_quantized}
\end{align}
We claim that the set of stable consensus fixed points of this system is exactly the set of stable fixed points of $f_Q$ itself, i.e., the points $e^*$ satisfying $(x - e^*)(f_Q(x) - x) < 0$ for all $x \neq e^*$ in a neighborhood of $e^*$.

\emph{Proof sketch.} At a perfect consensus $e_{i,t} = e^*$ for all $i$, \eqref{eq:consensus_quantized} gives $e^* = f_Q(e^*)$, so $e^*$ must be a fixed point of $f_Q$. For stability, consider an instantaneous small perturbation $e_{i,t} = e^* + \delta_{i,t}$ at time $t$. Since $h$ is a linear averaging function, the mean perturbation $\bar{\delta}_{t+1} = f_Q'(e^*)\,\bar{\delta}_t$ to leading order, where the derivative is understood in the sense of a local linear approximation. The consensus is stable when $|f_Q'(e^*)| < 1$, which corresponds precisely to $e^*$ being a stable fixed point of $f_Q$. For the standard quantization function with $f_Q(x) = q_j$ for $x \in (b_{j-1}, b_j]$, the fixed points are the quantization levels $q_j$ that lie within their own quantization bin (i.e., $q_j \in (b_{j-1}, b_j]$), and since $f_Q'(e^*) = 0$ at these points, they are always stable.

\subsection{Limited attention model: nonlinearity in aggregation and the singular $\omega \to \infty$ limit}

In the limited attention model, $f$ is the identity function and $h$ averages the top-$L$ (by absolute magnitude) messages. Since $h$ satisfies the internality property ($\min\{m_j\} \leq h \leq \max\{m_j\}$), every perfect consensus $e_{i,t} = e^*$ for all $i$ is trivially a fixed point. This means, when all messages are identical, top-$L$ selection has no effect and $h(e^*, \ldots, e^*) = e^*$. Therefore, the $\omega \to \infty$ limit yields a degenerate continuum of fixed points, has no bistability or multistability in the sense that it does not have coexisting attractors.

Despite this degeneracy, numerical simulations and mean-field results at finite $\omega$ reveal one or two isolated nonzero stable fixed points (Fig.~3G, Fig.~\ref{fig:BifurJaco_TopK}). We now show that these arise from a distributional mechanism that operates only when agents' estimates have nonzero variance, and that the $\omega \to \infty$ limit is singular.

\subsubsection{Scaling analysis} Using the mean-field map (Eqs.~\ref{eq:mf_topl_mu_update}--\ref{eq:mf_topl_sigma_update}), we analyze the steady state $(\mu^*, \sigma^*)$ as $\omega \to \infty$. Numerical results show that $z^* = \mu^*/\sigma^*$ grows as $O(\omega)$. We consider the two branches separately.

\emph{Branch with $z^* \to +\infty$.} In this regime, the distribution $\mathcal{N}(\mu^*, \sigma^{*2})$ is concentrated far from zero on the positive side. The top-$L$ selection threshold satisfies $q \approx \mu^* + c\,\sigma^*$ where $c = \Phi^{-1}(1 - L/k)$ is a constant, and nearly all selected messages come from the right tail. The conditional statistics satisfy
\begin{align}
    \mu_{\mathrm{sel}} - \mu^* &= +\sigma^* \cdot \frac{k\,\varphi(c)}{L} + O(\sigma^* \, e^{-z^{*2}/2}), \label{eq:musel_asymp_pos} \\
    \sigma_{\mathrm{sel}}^2 &= B \cdot \sigma^{*2} + O(\sigma^{*2} \, e^{-z^{*2}/2}), \label{eq:sigmasel_asymp_pos}
\end{align}
where $\varphi$ is the standard normal density and $B = 1 - c\,\varphi(c)\,k/L - (\varphi(c)\,k/L)^2$ is a positive constant satisfying $B < l$.

\emph{Branch with $z^* \to -\infty$.} By symmetry, the distribution is concentrated on the negative side, and the top-$L$ selection preferentially samples the left tail. The conditional mean shift has the opposite sign:
\begin{align}
    \mu_{\mathrm{sel}} - \mu^* &= -\sigma^* \cdot \frac{k\,\varphi(c)}{L} + O(\sigma^* \, e^{-z^{*2}/2}), \label{eq:musel_asymp_neg}
\end{align}
while $\sigma_{\mathrm{sel}}^2$ has the same leading-order expression as \eqref{eq:sigmasel_asymp_pos}.

The following part of the analysis is identical for both branches up to a sign. Substituting into the $\sigma^*$ fixed-point equation and retaining leading-order terms for large $\omega$:
\begin{align}
    \sigma^{*2} = \frac{\sigma_n^2}{(1 + L\omega)^2 - L\omega^2 B} \approx \frac{\sigma_n^2}{L(L - B)\,\omega^2}, \label{eq:sigma_scaling}
\end{align}
confirming $\sigma^* = O(1/\omega)$. Substituting into the $\mu^*$ fixed-point equation:
\begin{align}
    \mu^* &= \frac{G}{1 + L\omega} + \frac{L\omega}{1 + L\omega}\left(\mu^* \pm \sigma^* \cdot \frac{k\,\varphi(c)}{L}\right),
\end{align}
where $+$ corresponds to the $z^* \to +\infty$ branch and $-$ to the $z^* \to -\infty$ branch. Solving:
\begin{align}
    \mu^* = G \pm L\omega \cdot \sigma^* \cdot \frac{k\,\varphi(c)}{L} = G \pm \omega\,\sigma^* \cdot k\,\varphi(c).
\end{align}
Since $\omega \cdot \sigma^*$ converges to the constant $\sigma_n / \sqrt{L(L-B)}$, we obtain
\begin{align}
    \mu^*_{\pm} \;\to\; G \;\pm\; \Delta, \qquad \Delta \;=\; \frac{k\,\varphi(c)}{\sqrt{L(L-B)}} \; \sigma_n, \qquad \text{as } \omega \to \infty. \label{eq:mu_limit}
\end{align}
The two fixed points are symmetric about $G$, so their sum equals $2G$, consistent with numerical observations. The fixed point $\mu^*_- = G - \Delta$ can have the opposite sign to $G$ when $\Delta > |G|$, which is the condition for collective lock-in.

\subsubsection{Interpretation} Each nonzero fixed point at finite $\omega$ is sustained by a balance between two vanishingly small effects: (i) the observational noise, which maintains a small but nonzero spread $\sigma^* = O(1/\omega)$ in agents' estimates, and (ii) the amplification bias of top-$L$ selection, which preferentially samples messages aligned with the mean and shifts $\mu_{\mathrm{sel}}$ away from $\mu^*$ by an amount $O(\sigma^*)$. The environmental driving term $G/(1 + L\omega)$ also vanishes at the same rate. These three forces balance to produce an isolated fixed point at a specific value of $\mu^*$ that converges to a finite limit (\eqref{eq:mu_limit}).

Crucially, this mechanism is invisible in the naive $\omega \to \infty$ analysis where one sets $\sigma^* = 0$ and obtains a degenerate continuum of consensus fixed points. The correct asymptotic structure is that of a \emph{singular perturbation}: the observational noise $\sigma_n$, although its direct contribution to agents' estimates vanishes as $\omega \to \infty$, acts as a singular perturbation that selects discrete fixed points from the degenerate continuum. 

\subsubsection{Contrast with the quantized message model} The two models thus exhibit fundamentally different mechanisms of collective lock-in. In the quantized message model, the lock-in states are determined by the encoding nonlinearity $f_Q$ and exist as stable fixed points at $\omega \to \infty$. In the limited attention model, the encoding is linear ($f = \mathrm{id}$), the aggregation nonlinearity produces a degenerate landscape in the consensus limit, and the specific lock-in states are selected by a singular perturbation involving the interplay of observational noise, in-degree, and the attention window $L$. Despite these different origins, both mechanisms produce qualitatively similar phenomenology at finite $\omega$: multistability, hysteresis, and echo chamber-like locked-in states.

\subsection{Connecting collective decision-making and consensus formation} \label{sec:consensus}

It is tempting to understand the emergence of collective lock-in by examining the pure consensus limit $\omega \to \infty$, where personal observations play no role. In this limit, if the consensus-forming dynamics admits a nonzero steady state $e_{\mathrm{ss}} \neq 0$, symmetry guarantees a corresponding state $-e_{\mathrm{ss}}$. Since the system possesses only a single steady state $e_{\mathrm{ss}} = G_t$ when $\omega = 0$, one naturally expects bifurcations to create these additional steady states at some finite $\omega$, leading to collective lock-in. This reasoning applies cleanly to the quantized message model, where the message set $Q$ directly provides bounded nonzero consensus states in the $\omega \to \infty$ limit. However, this framing does not generalize to all models. In the limited attention model, the $\omega \to \infty$ limit is singular: because both the environmental driving force and the variance of agents' estimates vanish as $O(1/\omega)$, the landscape of steady states flattens and every consensus becomes a fixed point---a degenerate scenario that provides no information about which specific distortion states exist at finite $\omega$. The discrete nonzero fixed points observed at finite $\omega$ in this model arise from a delicate balance between vanishingly small noise and the amplification bias of top-$L$ selection, a balance that is invisible in the naive $\omega \to \infty$ limit. This highlights that collective lock-in can emerge through mechanisms that are not captured by analyzing pure consensus-forming dynamics alone.

\begin{figure}
    \centering
    \includegraphics[width=0.75\linewidth]{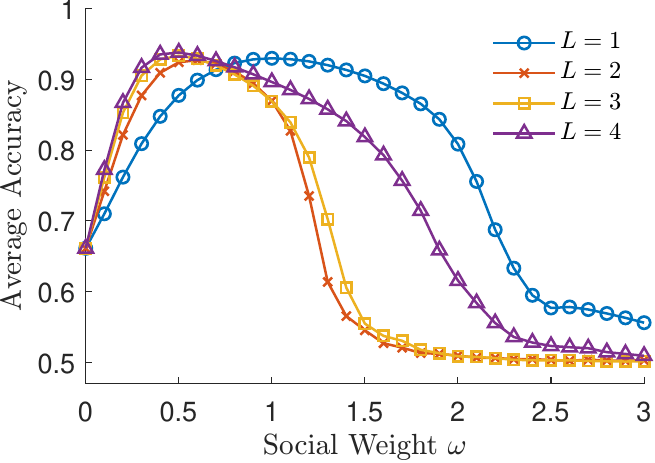}
    \caption{Average accuracy as a function of social weight for the limited attention model with different $L$. Across all values of $L$, a sudden drop in accuracy is observed at high social weight. In all plots, we use ER networks with $N=512$ and $\langle k\rangle=8$.}
    \label{fig:SI_TopKm_L}
\end{figure}

\section{Limited attention model with competition between social messages and personal observations} \label{sec:SI_TopOM}

\begin{figure}
    \centering
    \includegraphics[width=0.75\linewidth]{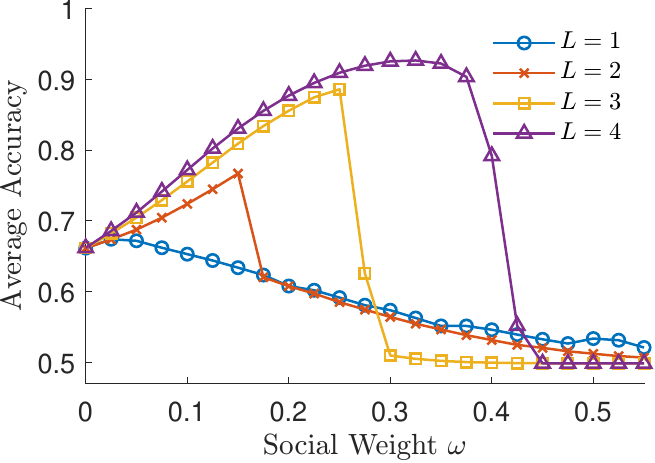}
    \caption{Average accuracy as a function of social weight for the limited attention model in which personal observations compete with social messages in the selection pool, with $L=1,2,3,4$. Other than $L=1$, a sudden drop in accuracy is observed at high social weight. In all plots, we use ER networks with $N=512$ and $\langle k\rangle=8$.}
    \label{fig:SI_TopK}
\end{figure}

In the main text, when discussing the limited attention model, we focus on the scenario in which each agent selects a limited number of social messages while always incorporating its personal observation. Here, we consider an alternative scenario in which personal observations are included in the same selection pool as social messages, and only the top-$L$ inputs are retained. In this variant of the limited attention model, personal observations may be ignored if they are weaker than the top $L$ social messages.
The results are shown in Fig.~\ref{fig:SI_TopK}, where a similar sudden drop in average accuracy is observed. However, the mean-field analysis for this variant appears to be more challenging than in the baseline scenario and we reserve this analysis for future work.

\section{Effects of including individual memory} \label{sec:SI_individualmemory}
\begin{figure}
    \centering
    \includegraphics[width=\linewidth]{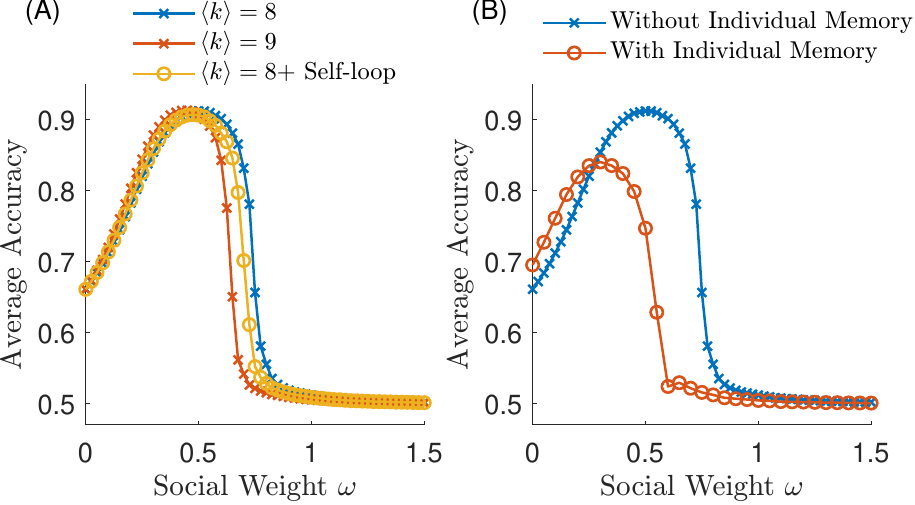}
    \caption{Effects of individual memory. 
    (A) Results when adding self-loops to all agents. In this case, adding one self-loop to each agent in an initially $\langle k \rangle=8$ ER network shifts the performance curve to a position between the $\langle k \rangle=8$ and $\langle k \rangle=9$ curves. 
    (B) Results with a Bayesian update rule that uses the previous estimate $e_{i,t-1}$ as a prior, compared with the baseline model without such individual memory.}
    \label{fig:selfloop}
\end{figure}

We test two ways of incorporating individual memory into the system.
The first requires only a minimal modification of the main-text model by adding a self-loop to each agent, so that an agent's previous estimate influences its future decision in a manner analogous to a neighboring agent's message.

The second introduces a pseudo-Bayesian update, inspired by the non-Bayesian framework of Ref.~\cite{jadbabaie2012non}. We refer to this update as pseudo-Bayesian because Bayes' rule is used to incorporate each agent's new personal observation, but the subsequent combination with social messages is not fully Bayesian. In particular, agents are not assumed to know the true likelihood function generated by their neighbors' messages, nor the network topology or updating rules of their neighbors. Instead, each agent uses its previous estimate $e_{i,t-1}$ as a prior belief and updates it using its new personal observation $o_{i,t}$. The resulting posterior-like belief is then combined with social messages through a convex combination.
During the personal update step, the previous estimate $e_{i,t-1}$ is treated as a log-likelihood ratio, as in the original model of Ref.~\cite{torney2015social}. The update rules are therefore
\begin{align}
    b_{i,t}^{\mathrm{prior}} &= e_{i,t-1}, \\
    b_{i,t}^{\mathrm{post}} &= b_{i,t}^{\mathrm{prior}} + \Delta b(o_{i,t}) \notag\\
    &= e_{i,t-1} + \log \frac{P(o_{i,t}\mid G_t=+1)}{P(o_{i,t}\mid G_t=-1)} \notag\\
    &= e_{i,t-1} + \frac{2}{\sigma_n^2} o_{i,t}, \\
    e_{i,t} &= \frac{1}{1+k_i\omega'}\, b_{i,t}^{\mathrm{post}} + \frac{\omega'}{1+k_i\omega'} \sum_{j\in N_i} m_{j,t-1}.
\end{align}
Here $\omega'$ plays a role similar to the social weight $\omega$ as in the models in the main text. The social weight $\omega$ can be interpreted as the relative weight of each piece of social message $m$ compared to the personal observation $o$. Here, $\omega'$ is instead the relative weight of each social message $m$ compared to the updated personal belief $b_{i,t}^{\mathrm{post}}$ rather than directly to the raw personal observation $o_{i,t}$. Thus, we have $\omega = \sigma_n^2\omega'/2$.

A result following the simplest extension, where a self-loop is added to all agents in a network, is shown in Fig.~\ref{fig:selfloop} (A). With an ER network with average in-degree $k=8$ as a baseline control group, we add one self-loop to each agent. The agents treat this self-loop the same way as a link from a neighbor. The result in Fig.~\ref{fig:selfloop} (A) shows that the resulting performance curve of the self-loop network is somewhere in between the $k=8$ and $k=9$ curves. The optimal global accuracy level is slightly lower than both the $k=8$ or $k=9$ cases. These results suggest that, while this self-loop adds one more message source so that the behavior is more similar to the $k=9$ case, this self-loop link still functions differently from the normal social link, which can provide information that is more independent.

A result with a Bayesian update rule is shown in Fig.~\ref{fig:selfloop} (B). 
The overall shape of the curve remains qualitatively similar to that of the baseline model without individual memory. However, the maximum achievable accuracy is lower when individual memory is present, and the sharp decline in accuracy occurs at a smaller value of the social weight $\omega$. Thus, introducing individual memory does not qualitatively alter the collective phenomena studied here. The transition to collective lock-in persists, while the overall performance becomes slightly worse. For clarity, we therefore focus on the simpler baseline model in the main text. This modeling choice is primarily methodological rather than biological. Individual memory may be important in real systems, but the goal here is to isolate the effects of social communication mechanisms.




\section{Heterogeneous networks can weaken collective lock-in; acyclic networks eliminate it} \label{sec:SI_heter}

\begin{figure}
    \centering
    \includegraphics[width=0.75\linewidth]{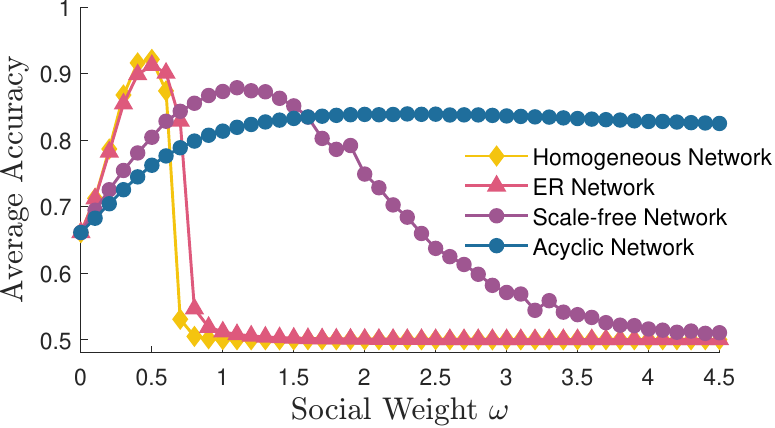}
    \caption{Average decision accuracy as a function of social weight in different network topologies with the same mean in-degree in the binary message model. Among homogeneous, ER, and scale-free networks, increasing heterogeneity in the in-degree distribution introduces more low-degree agents, which lowers the peak accuracy but delays and softens the accuracy collapse. The acyclic network has the lowest peak accuracy but does not exhibit a sudden drop caused by hysteresis or locked-in dynamics.}
    \label{fig:SI_heterNets}
\end{figure}

Given the results in the main text that agents with lower in-degree appear more sensitive to environmental shifts (Fig.~5 A, B), we asked whether increasing the proportion of such agents, while keeping all other factors fixed, could enhance the overall sensitivity of collective decision-making. To test this idea, we fix the average degree of the communication network and compare three network configurations that differ only in their in-degree distributions. In this way, the total number of social links is kept constant, but these links are distributed differently across agents, resulting in different proportions of low in-degree individuals.
The three network types we consider are: a homogeneous random network, where all agents have identical in-degree $k$; an ER network, where in-degrees follow a binomial distribution; and a scale-free network, where in-degrees follow a power-law distribution and heterogeneity is the strongest.
As shown in Fig.~\ref{fig:SI_heterNets}, networks with lower in-degree heterogeneity achieve higher optimal decision accuracy but become locked-in at smaller values of the social weight $\omega$. This indicates that lower heterogeneity makes the system more robust but less sensitive in the robustness–sensitivity trade-off.

Figure ~\ref{fig:SI_heterNets} further shows that acyclic network structures can eliminate locked-in states and hysteresis, but at the cost of reduced robustness, resulting in a lower optimal accuracy than in the other three network classes. In an acyclic network, information propagates in a strictly feedforward manner without forming loops. This feedforward, hierarchical organization ensures that outdated social information is not perpetuated within the network in a way that could influence future decisions. However, the same lack of recurrent reinforcement also reduces the collective’s ability to average out observational noise. Decisions are thus more sensitive to stochastic fluctuations, especially when these fluctuations originate from upstream agents. 

\section{Alternative communication network structures} \label{sec:SI_addnets}

\begin{figure}
    \centering
    \includegraphics[width=\linewidth]{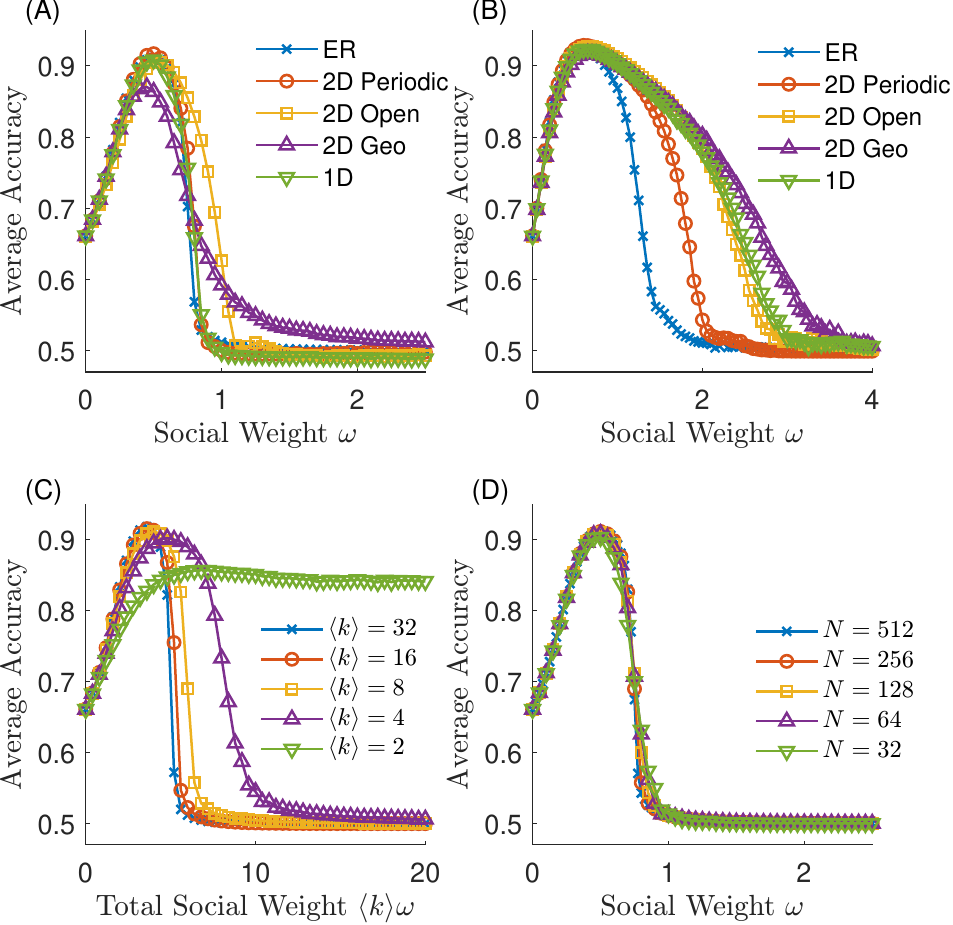}
    \caption{Effects of different communication networks. 
    (A) Results for different network topologies with the binary message model. $N=256$, $\langle k\rangle=8$. 
    (B) Results for different network topologies with the limited attention model. $L=2$, $N=256$, $\langle k\rangle=8$. 
    (C) Results for different average in-degree in the ER network with binary messages. $N=256$. 
    (D) Results for different network sizes in the ER network with binary messages. $\langle k\rangle=8$.
    In all plots, we use $\sigma_n=2.24$.}
    \label{fig:addNetwork}
\end{figure}

\begin{figure}
    \centering
    \includegraphics[width=0.85\linewidth]{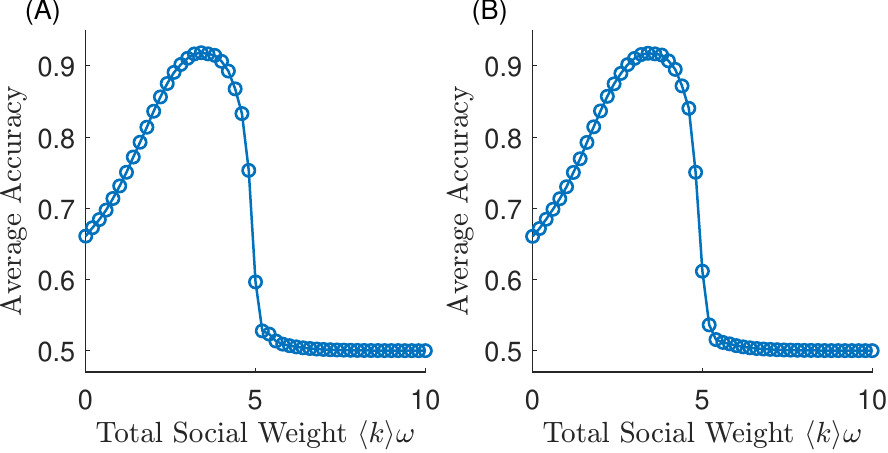}
    \caption{Accuracy of collective decision-making in (A) complete graph and (B) complete bipartite graph. Both plots are generated using $N=128$, repeated over 50 trials.}
    \label{fig:addCompleteNetwork}
\end{figure}

In the main text, most of our analyses focused on random communication networks. In this supplementary section, we test the robustness of our results under network structures that are more regular or exhibit stronger spatial organization. Specifically, we consider the following four types of networks:
(i) 2-D lattice network with periodic boundary conditions,
(ii) 2-D lattice network with open boundary conditions,
(iii) 2-D random geometric graph, where nodes are randomly distributed in the plane and nodes within a fixed radius $r_c$ are connected,
(iv) 1-D lattice network with periodic boundary conditions.
In Fig.~\ref{fig:addNetwork}(A, B), we show the results of collective decision-making with the binary message model and the limited attention model on these different network topologies. Across these network structures, we observe a qualitatively similar phenomenon: a sudden drop in average accuracy within a certain range of social weight $\omega$.

In Fig.~\ref{fig:addNetwork}(C, D), we further examine the ER network with different average in-degrees $\langle k \rangle$ and network sizes $N$. When $\langle k \rangle$ is very small, consistent with our discussion in the main text that low in-degree can break locked-in states, we do not observe a sudden decrease in accuracy. Otherwise, varying $\langle k \rangle$ and $N$ mainly produces quantitative changes in the shape of the curve without altering the overall qualitative behavior.

In the main text, we briefly mentioned a previous study on opinion dynamics with quantized communicated behaviors, which showed that quantization can generate nonconsensus equilibria and large deviations from consensus on general graphs~\cite{ceragioli2018consensus}. In particular, that work proved that the distance from consensus can become very large for some network topologies, while also showing that, for their model, all trajectories converge to consensus on complete graphs and complete bipartite graphs~\cite{ceragioli2018consensus}.
By contrast, in our framework, collective lock-in persists even on complete graphs and complete bipartite graphs, as shown in Fig.~\ref{fig:addCompleteNetwork}. This suggests that the phenomenon studied here is mechanistically different from the topology-dependent nonconsensus effect reported in Ref.~\cite{ceragioli2018consensus}. In their case, the main effect is the emergence of disagreement or nonconsensus induced by quantized communicated states (without the issue of tracking a changing environment), whereas in our case, the dominant effect is a breakdown of environment tracking that can survive even under maximally mixed interaction topologies.

\section{Alternative environmental and noise settings} \label{sec:SI_EnvNoise}

\begin{figure}
    \centering
    \includegraphics[width=\linewidth]{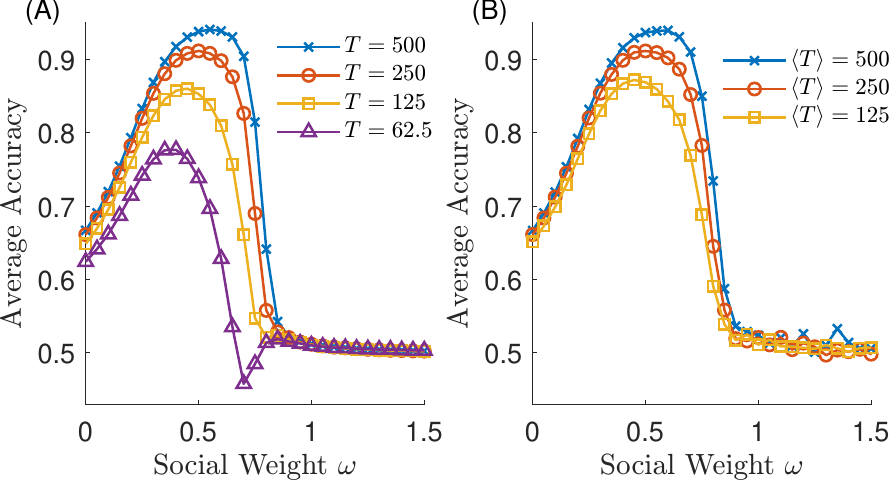}
    \caption{Effects of different environmental dynamics $G_t$. (A) Results for periodic environments with different periods $T$ in the binary message model. (B) Results for a randomly switching environment, where the waiting time between flips follows an exponential distribution. In all plots, $N=256$ and $\langle k\rangle=8$. }
    \label{fig:addEnv}
\end{figure}

\begin{figure}
    \centering
    \includegraphics[width=0.85\linewidth]{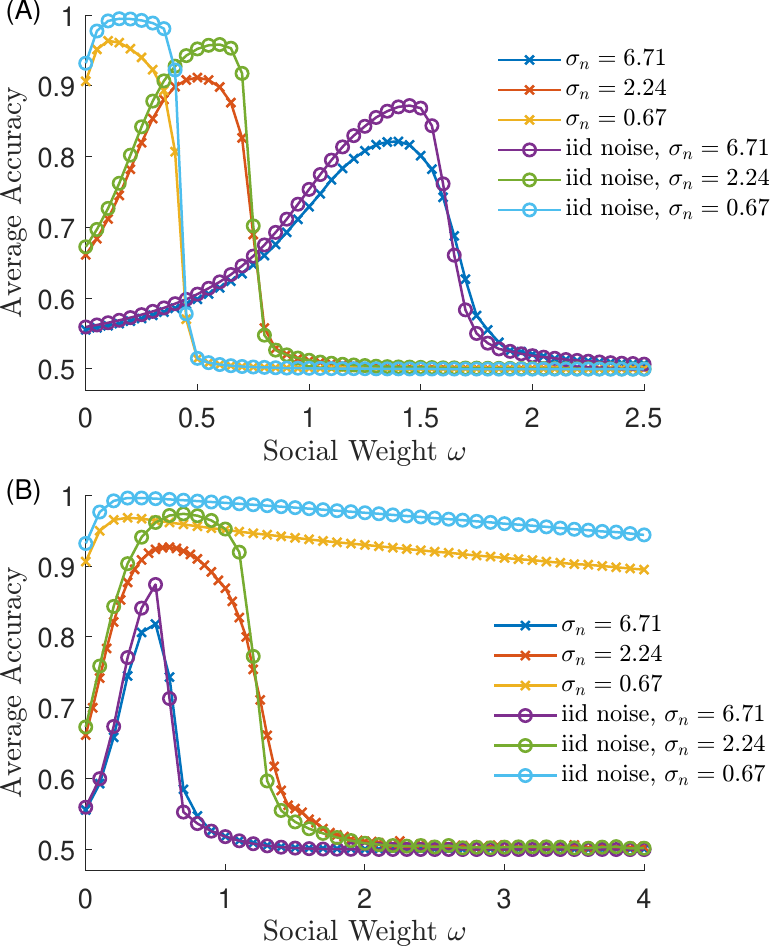}
    \caption{Effects of different levels and types of observational noise. We test both different noise amplitudes and two noise generation mechanisms: noise generated by an Ornstein--Uhlenbeck process, which introduces temporal autocorrelation, and i.i.d.\ Gaussian noise. (A) Results for the binary message model. (B) Results for the limited attention model with $L=2$, where the bifurcation to bistability and the sudden drop in accuracy occur only when the noise level exceeds a certain threshold. In all plots, we use $N=256$ and $\langle k\rangle=8$.}
    \label{fig:addNoise}
\end{figure}

Here we test several settings beyond those considered in the main text. In the main text, we focused on a periodically switching environment in which the environmental state $G_t$ flips every $T=250$ time steps, with observational noise generated by an Ornstein--Uhlenbeck (OU) process,
\begin{align}
    do_t = -\lambda_g o_t\,dt + \sigma\,dW_t ,
\end{align}
which introduces temporal correlations in the observations. We used $\lambda_g=0.1$ and $\sigma=1.0$ in the main text, which results in $\sigma_n=2.24$.

In this section, we test additional environmental dynamics and noise models beyond those considered in the main text. We first examine different flipping frequencies by varying the period length $T$. As shown in Fig.~\ref{fig:addEnv}(A), more frequent environmental changes generally reduce decision accuracy. When switching becomes sufficiently fast, the average accuracy can even fall below $0.5$, reflecting the lag in the collective response of the agents to environmental changes.
Next, we consider environments in which the switching times are stochastic, with the waiting time between two flips following an exponential distribution, corresponding to a Poisson switching process. This introduces variability in the duration of each environmental state compared with the periodic switching considered in the main text. As shown in Fig.~\ref{fig:addEnv}(B), this stochasticity does not substantially alter the overall results of collective decision-making.

We further examine the effects of different levels and types of observational noise. The results are shown in Fig.~\ref{fig:addNoise}. Using i.i.d.\ observational noise instead of the temporally correlated Ornstein--Uhlenbeck noise slightly increases the average accuracy quantitatively, especially around the peak regions, but does not introduce qualitative changes to the shape of the curves. In the limited attention model, as shown in Fig.~\ref{fig:network}(B), the bifurcation to bistability and the associated sudden drop in accuracy occur only when the noise level exceeds a certain threshold.

An important implication of the results in Fig.~\ref{fig:addEnv} and Fig.~\ref{fig:addNoise} is that the ranges of social weight $\omega$ that optimize collective performance, versus those that produce lock-in, depend strongly on environmental conditions such as noise level and flipping frequency. In other words, there is no universally optimal level of social influence: a value of $\omega$ that is adaptive in one environment may become maladaptive in another. As a result, the same group of agents can switch from near-optimal tracking to severe lock-in solely due to environmental changes. For example, in Fig.~\ref{fig:addNoise}(A), the value of $\omega$ that maximizes performance when $\sigma_n=0.67$ falls within the lock-in regime when $\sigma_n=2.24$, where agents become almost completely insensitive to environmental change. More broadly, this implies that collective sensitivity to environmental dynamics depends not only on the interaction rules themselves, but on how well those rules are matched to the current environment.

\section{An ant trail recruitment model} \label{sec:SI_ant}
We consider an ant trail recruitment model inspired by pheromone-mediated foraging in ant colonies. Ant foraging trails are a classic example of self-organization, in which colony-level trail patterns emerge from local interactions mediated by pheromone deposition and trail following \cite{deneubourg1990self,beckers1993modulation,perna2012individual}. Motivated by these studies, we model social communication as indirect signaling through shared pheromone fields rather than direct message passing between agents.

\subsection{Model}

In our setting, each ant in the group must choose between two trails, $A$ and $B$, and attempts to select the one that leads to food. We assume that the food-bearing trail switches over time. At each time step, each agent receives a noisy private cue about the ground-truth environmental state of the food resource (denoted by $g_i(t)$), while also sensing the shared pheromone fields on the two trails left by previous agents (denoted by $\phi_A(t)$ and $\phi_B(t)$). The agent then chooses between two discrete actions corresponding to the two candidate trails. In the next time step, the ant is assumed to have completed a round trip between the nest and the selected trail, and then returns to the nest to participate in the next round of decision-making. 

Following classic ant trail recruitment models, the probability that agent $i$ chooses option $A$ is given by a nonlinear pheromone response function~\cite{deneubourg1990self,beckers1993modulation}
\begin{align}
P_i(A,t) =
\frac{(k_{\mathrm{ant}}+\mathrm{eff}_A^{(i)}(t))^{a}}
{(k_{\mathrm{ant}}+\mathrm{eff}_A^{(i)}(t))^{a} + (k_{\mathrm{ant}}+\mathrm{eff}_B^{(i)}(t))^{a}},
\end{align}
where $\mathrm{eff}_A^{(i)}(t)$ and $\mathrm{eff}_B^{(i)}(t)$ are the effective support for the two options, $k_{\mathrm{ant}}$ and $a$ are constants, and $P_i(B,t)=1-P_i(A,t)$. The agent's action $u_i(t) \in \{-1,1\}$ is then sampled according to these probabilities, where $u_i(t)=+1$ means agent $i$ picks trial $A$ at time $t$, and $u_i(t)=-1$ means agent $i$ picks trial $B$ at time $t$.
When deciding between the two trials, the effective support for the two options combines the private cue (e.g. individual memory or direct observation) and the pheromone signal,
\begin{align}
\mathrm{eff}_A^{(i)}(t) &= \frac{1}{1+\omega} \, \max(g_i(t),0) + \frac{\omega}{1+\omega} \phi_A(t),\\
\mathrm{eff}_B^{(i)}(t) &= \frac{1}{1+\omega} \, \max(-g_i(t),0) + \frac{\omega}{1+\omega} \phi_B(t),
\end{align}
where $\omega$ is the social weight.

After making a decision, the agent deposits pheromone on the selected option. Let $q_i(t)$ denote the amount of pheromone deposited by agent $i$. The pheromone levels evolve as
\begin{align}
\phi_A(t+1) &= (1-\rho_{\mathrm{ant}})\phi_A(t)
+ \sum_{i: u_i(t)=+1} q_i(t), \\
\phi_B(t+1) &= (1-\rho_{\mathrm{ant}})\phi_B(t)
+ \sum_{i: u_i(t)=-1} q_i(t),
\end{align}
where $\rho$ represents pheromone evaporation.
Results in Ref.~\cite{beckers1993modulation,jackson2007modulation} suggest that food quality and foraging outcome can modulate pheromone reinforcement. For Pharaoh’s ant, this modulation is through differences in marking intensity with a constant frequency of marking~\cite{jackson2007modulation}. For \textit{Lasius niger}, this modulation is through an all-or-nothing response~\cite{jackson2007modulation}. Individuals will lay trail pheromone only if the quality of a food source is greater than their personal threshold~\cite{mailleux2000ants}, which can vary among individuals. We therefore assume that the deposited pheromone $q_i(t)$ takes, in average, two different values depending on whether the agent's decision matches the environmental state,
\[
q_i(t)=
\begin{cases}
\eta_\mathrm{good} q_{\mathrm{good}}, & u_i(t)=G(t),\\
\eta_\mathrm{bad}q_{\mathrm{bad}}, & u_i(t)\neq G(t),
\end{cases}
\]
where $\eta_\mathrm{good}$ and $\eta_\mathrm{bad}$ represent the rates of ants marking the trial they pick, and if they do, $q_{\mathrm{good}}$ and $q_{\mathrm{bad}}$ represent the intensity of the marking. We have $\eta_\mathrm{good}q_{\mathrm{good}}>\eta_\mathrm{bad}q_{\mathrm{bad}}$.
The settings of the environmental state $G_t$ and the personal observations $g_{i,t}$ remain the same as the models in the main text. One does not need to pick a communication network because all the messages are transmitted through the pheromone fields.

Following Ref.~\cite{beckers1993modulation}, we use parameter values $k_{\mathrm{ant}}=6$ and $a=2$. Following Ref.~\cite{jackson2007modulation}, we use $q_{\mathrm{good}}=1.0$, $q_{\mathrm{bad}}=0.8$, $\eta_{\mathrm{good}}=\eta_{\mathrm{bad}}=0.4$ with the settings of Pharaoh’s ant. If we instead follow the settings of Lasius niger, then we can use $q_{\mathrm{good}}=q_{\mathrm{bad}}=1.0$, $\eta_{\mathrm{good}}=0.8$, and $\eta_{\mathrm{bad}}=0.2$. These changes in $q_i$ would still produce a similar curve of average accuracy versus social weight as in Fig.4 (B) and preserve all the phenomena, though the x-axis of social weight would be rescaled. We use $\rho_{\mathrm{ant}}=0.5$ and $N=50$. 

\subsection{Results}
The main results are shown in Fig. 4(A) in the main text. As $\omega$ increases, the  model undergoes a transition to a lock-in state and we observe hysteresis in the system.

\section{Multi-animal directional collective decision-making with ring attractor neural networks} \label{sec:SI_ring}

As an additional model that implements a (soft) winner-take-all (WTA) mechanism, we consider a ring attractor neural network model originally developed in neuroscience~\cite{skaggs1994model,zhang1996representation}. Ring attractor networks are recurrent neural networks that support a continuous attractor manifold with circular topology, on which a localized bump of activity is maintained~\cite{chaudhuri2016computational}. This activity bump encodes a circular variable, such as an animal’s internal estimate of heading direction. 

Such models have been used to describe neural representations of directional variables and have also been proposed as mechanisms for integrating directional information through soft WTA dynamics~\cite{maass2000computational,sridhar2021geometry,sayin2025behavioral,gorbonos2024geometrical,oscar2023simple}. Here, we use this class of models as a representative example of a biologically motivated soft WTA aggregation mechanism.
Instead of assuming a black-box mechanism that directly produces a WTA outcome given the inputs, each animal agent is modeled with a ring attractor neural network that aggregates multiple directional inputs through its internal neural dynamics. In this sense, the ring attractor network provides a dynamical implementation of a nonlinear aggregation rule, in which competing directional cues interact through recurrent inhibition and local excitation.

The collective task we consider is inspired by the heading consensus problem~\cite{chate2008collective,li2008singularities}, where agents coordinate their movement directions through local interactions. In our setting, however, we consider a slightly different scenario. Instead of focusing purely on consensus formation, we assume that there exists a direction leading to a food resource, which may change over time. In addition to social interactions, as in the standard heading consensus problem, agents also receive personal observations of the environmental state and attempt to infer this underlying ground truth.

\subsection{Model}
The action space of each agent is directional, where $u \in (-\pi,\pi]$. The ground truth optimal direction to the food resource is presented by the variable $G_t \in (-\pi,\pi]$. To make things simpler and better aligned with other modes in this paper, assumed to always be in either direction $+\pi/2$ or $-\pi/2$. Thus, $G_t$ is still binary. At each time step, each agent has a personal observation ($o_{i,t}$) of the environment $G_t$, where we still have 
\begin{align}
    o_{i,t}=G_t+\sigma_n\zeta_{i,t},
\end{align}
where the term $\sigma_n\zeta_{i,t}$ is the observational noise. $\sigma_n$ is the coefficient controlling the magnitude of the noise.
Each agent can also perceive the moving direction of other agents, which is the social message in this model,
\begin{align}
    m_{i,t} = u_{i,t}.
\end{align}

As discussed above, each agent has its own ring attractor neural network to aggregate its personal observation and all the social messages it receives. Each of these neural networks contains $N_{\mathrm{ring}}=128$ neurons arranged on a circular topology, where neuron $m$ has a preferred direction $\phi_m = 2\pi m/N_{\mathrm{ring}}$. The neural activity vector $a_m$ represents the activity level of neurons tuned to different directions. Directional inputs are provided as a set of cues $(\theta_l,s_l)$, where $\theta_l$ is the direction of the $L$-th cue and $s_l$ is its strength. These cues include both the agent's personal observation and the directional messages received from neighboring agents. Each cue generates a localized external input on the ring through a Gaussian tuning curve centered at $\theta_l$.

Neural activities evolve through recurrent interactions on the ring attractor. The recurrent connectivity matrix $W$ is defined by a cosine-shaped interaction kernel that provides local excitation and long-range inhibition, enabling the network to sustain a single localized bump of activity. Specifically, the interaction strength between neurons with preferred directions $\phi_n$ and $\phi_m$ is given by
\begin{align}
W_{nm} = \cos\!\left(\pi \left|\frac{\phi_n - \phi_m}{\pi}\right|^{\nu}\right),
\end{align}
where $\nu$ controls the sharpness of the interaction kernel. This connectivity structure produces local excitation and longer-range inhibition along the ring, allowing the network to stabilize a single localized bump of activity.
At each iteration the activity is updated according to
\begin{align}
a(t+1) =
(1-\lambda)a(t)
+ \gamma W \tanh(a(t))
+ \eta_{\mathrm{ring}} I_{\mathrm{ring}} ,
\end{align}
where $W$ is the recurrent interaction matrix determined by the circular distance between neurons, $I_{\mathrm{ring}}$ is the external input vector generated from the directional cues, and $\lambda$, $\gamma$, and $\eta_{\mathrm{ring}}$ control the decay, recurrent interaction, and input strengths respectively. The dynamics are iterated until convergence.

The final activity profile forms a single bump on the ring. The agent's inferred direction is then obtained as the circular mean of the neural activities,
\begin{align}
u_i = e_i = \mathrm{atan2}\!\left(
\sum_k a_k \sin\phi_k,
\sum_k a_k \cos\phi_k
\right).
\end{align}

\subsection{Results}

In the ring attractor models, the coefficient $\nu$ controls how soft or hard the WTA is. When $\nu=1.0$, there is no WTA anymore and the neural network dynamics are equivalent to the average-taking model. As shown in Fig.~4 (B) middle panel, the accuracy versus $\omega$ curves for the $\nu=1.0$ ring attractor model do not have a significant drop in high $\omega$ regions. When $\nu\to0$, the ring attractor model approaches the hard WTA model, where the agent selects a single input (from social messages or its direct measurement) and aligns with it. In Fig.~4 (A) middle panel, we show that in ring attractor models with $\nu=0.5$, high $\omega$ can lead to hysteresis and a sudden drop of accuracy. In Fig.~4 (A) lower panel, we plot the hysteresis loop of a group of agents with ring attractor neural models.

\section{Collective decision-making in a cortical neural network} \label{sec:SI_neural}

We have shown that quantized messaging can produce lock-in and hysteresis in generic models of collective decision-making.  Here, we ask whether the same phenomena arise in a recurrent cortical circuit, where the ``agents'' are neurons and the ``messages'' are transmitted across synapses.  Chemical synaptic transmission discretizes the presynaptic membrane potential via the spike threshold and then activates slow NMDA receptors~\cite{brunel2000dynamics,wang2002probabilistic,dayan2005theoretical}, making the system analogous to the quantized-message model. Gap junctions, by contrast, transmit the full continuous membrane potential directly~\cite{connors2004electrical,bennett2004electrical} and do not have a quantization mechanism.  We constructed a biophysically realistic spiking network following Wang~\cite{wang2002probabilistic} to test whether chemical synapses produce collective lock-in while gap junctions do not.

\subsection{Model}
\subsubsection{Single-neuron dynamics}
Each neuron $i$ is modeled as a leaky integrate-and-fire (LIF) unit~\cite{brunel2000dynamics,wang2002probabilistic,dayan2005theoretical}, whose membrane potential $V_i$ evolves according to
\begin{align}\label{eq:lif}
  \tau_m \frac{dV_i}{dt}
    = -(V_i - V_L) + I_i^{\mathrm{ext}}(t) + I_i^{\mathrm{rec}}(t),
\end{align}
where $\tau_m = 20$~ms is the membrane time constant, $V_L = -70$~mV is the leak reversal potential. $I_i^{\mathrm{ext}}(t)$ is the external input current, which corresponds to the personal observation in other models in this paper.
$I_i^{\mathrm{rec}}(t)$ is the recurrent synaptic current, which corresponds to the social messages in other models. When $V_i$ reaches the spike threshold $V_{\mathrm{th}} = -50$~mV, neuron~$i$ emits a spike, $V_i$ is reset to $V_{\mathrm{reset}} = -55$~mV, and the neuron enters an absolute refractory period of duration $\tau_{\mathrm{ref}} = 2$~ms during which $V_i$ is frozen.  These parameter values lie within the standard range for cortical pyramidal neurons~\cite{brunel2000dynamics,dayan2005theoretical}.

A single LIF neuron is not bistable: in the absence of recurrent input ($I^{\mathrm{rec}} = 0$), \eqref{eq:lif} is a linear first-order ODE with a unique stable fixed point $V^* = V_L + I^{\mathrm{ext}}$. Any collective bistability that emerges in the network model must therefore arise from the recurrent interactions, not from single-cell intrinsic dynamics.

\subsubsection{External input}

Similar to other models in this paper, we assume a binary environment $G(t)$ that the neurons are trying to track. $I_i^{\mathrm{ext}}(t)$ is the feedforward sensory current received by agent $i$, which encodes the environment state. We assume
\begin{align}\label{eq:sens}
  I_i^{\mathrm{ext}}(t) = I_0 + \beta\, G(t) + \zeta_i(t),
\end{align}
where $I_0$ is a baseline drive (tuned so that the spontaneous firing rate is $\sim\!20$~Hz in the absence of stimulus and recurrent input), $\beta$ controls the stimulus strength, and $\zeta_i(t)$ is an Ornstein--Uhlenbeck noise process following the same setting as in other models with standard deviation $\sigma_n$.
The environment switches between $G = 1$ (stimulus present) and $G = 0$ (stimulus absent). Here, we use $G=0$ instead of $G=-1$ for the stimulus-absent state, which fits more naturally into the narrative. 
Crucially, the baseline drive $I_0$ is set \emph{below} the effective firing threshold $V_{\mathrm{th}} - V_L = 20$~mV.  In the absence of both stimulus and recurrent input, neurons fire at a low spontaneous rate (usually less than 3~Hz) driven solely by stochastic fluctuations in the sensory noise $\zeta_i(t)$. 

\subsubsection{Recurrent input: chemical synapses (spike-coded)}

For the chemical-synapse model, recurrent excitation is mediated by conductance-based synapses with both fast AMPA and slow NMDA receptor components, following the biophysically realistic framework of Ref.~\cite{wang2002probabilistic} and Ref.~\cite{brunel2000dynamics}.
Each neuron~$j$ maintains two synaptic gating variables.  The fast AMPA variable $s_j^{\mathrm{AMPA}}$ obeys simple exponential dynamics:
\begin{align}\label{eq:ampa}
  \frac{ds_j^{\mathrm{AMPA}}}{dt} = -\frac{s_j^{\mathrm{AMPA}}}{\tau_{\mathrm{AMPA}}},
  \qquad
  s_j^{\mathrm{AMPA}} \leftarrow s_j^{\mathrm{AMPA}} + 1 \;\;\text{upon spike},
\end{align}
with $\tau_{\mathrm{AMPA}} = 2$~ms.

The slow NMDA gating variable $s_j^{\mathrm{NMDA}}$ is described by a two-variable model~\cite{wang2002probabilistic}:
\begin{align}
  \frac{dx_j}{dt} &= -\frac{x_j}{\tau_{\mathrm{rise}}},
  \qquad
  x_j \leftarrow x_j + 1 \;\;\text{upon spike},
  \label{eq:nmda_x}
  \\[4pt]
  \frac{ds_j^{\mathrm{NMDA}}}{dt}
  &= -\frac{s_j^{\mathrm{NMDA}}}{\tau_{\mathrm{NMDA}}}
     + \alpha\, x_j \bigl(1 - s_j^{\mathrm{NMDA}}\bigr),
  \label{eq:nmda_s}
\end{align}
where $\tau_{\mathrm{rise}} = 2$~ms is the NMDA rise time constant, $\tau_{\mathrm{NMDA}} = 100$~ms is the decay time constant, and $\alpha = 0.5$~ms$^{-1}$ controls the saturation rate~\cite{wang2002probabilistic}.  The factor $(1 - s_j^{\mathrm{NMDA}})$ in Eq.~\ref{eq:nmda_s} enforces biophysical saturation, bounding $s_j^{\mathrm{NMDA}} \in [0, 1]$.  At steady state with constant firing rate~$r$, the NMDA gating variable converges to
\begin{align}\label{eq:nmda_ss}
  s_\infty(r)
    = \frac{\alpha\, r\, \tau_{\mathrm{rise}}\, \tau_{\mathrm{NMDA}}}
           {1 + \alpha\, r\, \tau_{\mathrm{rise}}\, \tau_{\mathrm{NMDA}}},
\end{align}
yielding $s_\infty \approx 0.20$ at $r = 5$~Hz and $s_\infty \approx 0.75$ at $r = 30$~Hz.

The recurrent current to neuron~$i$ is the sum of AMPA and NMDA components, both conductance-based with excitatory reversal potential $V_E = 0$~mV:
\begin{align}\label{eq:Irec_chem}
  I_i^{\mathrm{rec}} = \omega \Bigl[
    g_{\mathrm{AMPA}}\,(V_E - V_i)\,\bar{s}_i^{\mathrm{AMPA}}
    \;+\;
    g_{\mathrm{NMDA}}\,(V_E - V_i)\,B(V_i)\,\bar{s}_i^{\mathrm{NMDA}}
  \Bigr],
\end{align}
where $\omega$ is the recurrent coupling strength (the neural analogue of social weight), $g_{\mathrm{AMPA}}$ and $g_{\mathrm{NMDA}}$ are the synaptic conductance ratios (normalized by the leak conductance $g_L$, so that $I^{\mathrm{rec}}$ has units of mV in ~\eqref{eq:lif}), and
\begin{align}
  \bar{s}_i^{X} = \frac{1}{k_i}\sum_{j} A_{ij}\, s_j^{X},
  \qquad X \in \{\mathrm{AMPA},\, \mathrm{NMDA}\},
\end{align}
are the neighbor-averaged gating variables.  The function
\begin{align}\label{eq:mgblock}
  B(V) = \frac{1}{1 + [\mathrm{Mg}^{2+}]\,\exp(-0.062\,V)\,/\,3.57}
\end{align}
is the voltage-dependent Mg$^{2+}$ block of NMDA receptors, with $[\mathrm{Mg}^{2+}] = 1$~mM.

Because the LIF membrane potential is confined to the interval $[V_{\mathrm{reset}},\, V_{\mathrm{th}}] = [-55,\, -50]$~mV, the Mg$^{2+}$ block factor $B(V)$ varies only between $\approx 0.11$ and $\approx 0.14$ and is effectively constant.  Thus, in a single-compartment LIF model, $B(V)$ does not contribute significant voltage-dependent nonlinearity; we retain it for biophysical completeness but note that the nonlinearity responsible for bistability arises from a different source.
We emphasize that, unlike simplified models in which a sigmoidal function is applied to the presynaptic activity by construction, the present model contains no hand-tuned nonlinearity.  Instead, the effective sigmoidal transfer function arises as an emergent property of the LIF neuron operating in the fluctuation-driven regime~\cite{brunel2000dynamics}.  In this regime, the neuron's mean firing rate as a function of its total input $I$ is given by
\begin{align}\label{eq:fIcurve}
  \frac{1}{r}
    = \tau_{\mathrm{ref}}
    + \tau_m\sqrt{\pi}
      \int_{(V_{\mathrm{reset}} - I)/\sigma}^{(V_{\mathrm{th}} - I)/\sigma}
        e^{u^2}\bigl(1 + \mathrm{erf}(u)\bigr)\,du,
\end{align}
where $\sigma$ is the standard deviation of the total input fluctuations~\cite{brunel2000dynamics}.  This function $r = \Phi(I)$ has a sigmoidal shape: it is vanishingly small when $I$ is well below threshold ($I \ll V_{\mathrm{th}} - V_L$), rises steeply near threshold, and saturates at high input due to the refractory period.  The steepness of this emergent sigmoid is controlled by the noise amplitude~$\sigma$: smaller noise produces a steeper sigmoid (approaching a hard threshold in the deterministic limit $\sigma \to 0$), while larger noise produces a shallower one.

The combination of the emergent sigmoidal transfer function and positive recurrent feedback creates bistability.  In steady state, the self-consistent condition for the population firing rate~$r$ is
\begin{align}\label{eq:selfconsist}
  r = \Phi\!\Bigl(I_0 + \omega\, g_{\mathrm{eff}}\, s_\infty(r)\Bigr),
\end{align}
where $g_{\mathrm{eff}} = g_{\mathrm{NMDA}}\,(V_E - \bar{V})\,B(\bar{V})$ is the effective recurrent conductance (approximately constant for typical membrane potentials $\bar{V} \in [V_{\mathrm{reset}}, V_{\mathrm{th}}]$), and $s_\infty(r)$ is given by Eq.~\ref{eq:nmda_ss}.  At low recurrent coupling~$\omega$, this equation has a unique solution (a low-firing state).  As $\omega$ increases, the positive feedback through the NMDA-mediated recurrent loop becomes strong enough for Eq.~\ref{eq:selfconsist} to admit three solutions: a stable low-firing state, an unstable intermediate state, and a stable high-firing state.  The low-firing state corresponds to the fluctuation-driven regime where $I_0 + \omega\, g_{\mathrm{eff}}\, s_\infty(r_{\mathrm{low}})$ remains below threshold; the high-firing state is self-sustained because $I_0 + \omega\, g_{\mathrm{eff}}\, s_\infty(r_{\mathrm{high}})$ exceeds threshold, even in the absence of stimulus.

\subsubsection{Recurrent input: gap junctions (voltage-coded)}
For the gap-junction model, the recurrent current is proportional to the difference between the average membrane potential of neuron~$i$'s neighbors and its own membrane potential~\cite{connors2004electrical,bennett2004electrical}:
\begin{align}\label{eq:gap}
  I_i^{\mathrm{rec}} = \omega\, g_{\mathrm{gap}} \Bigl(
    \frac{1}{k_i}\sum_{j} A_{ij}\, V_j - V_i
  \Bigr).
\end{align}
This current vanishes when all connected neurons have similar membrane potentials and therefore cannot provide net excitatory drive to sustain elevated activity after stimulus removal. 

\subsubsection{Readout}
The collective state of the network is characterized by the population-averaged instantaneous firing rate $\bar{r}(t)$, estimated by passing each neuron's spike train through an exponential filter with time constant $\tau_r$:
\begin{align}
  \tau_r \frac{d\hat{r}_i}{dt} = -\hat{r}_i, \qquad
  \hat{r}_i \leftarrow \hat{r}_i + 1 \;\;\text{upon spike},
  \qquad
  \bar{r}(t) = \frac{1}{N}\sum_i \hat{r}_i(t) \cdot \frac{1000}{\tau_r},
\end{align}
where the factor $1000/\tau_r$ converts to units of Hz.  The readout filter does not participate in the network dynamics.

\subsection{Results}
The main results are shown in Fig. 4(C) in the main text. As $\omega$ increases, the chemical-synapse (spike train) model undergoes an abrupt transition.  Beyond a critical~$\omega$, the network locks into a persistent high-firing state after stimulus removal, a neural analogue of the echo chambers identified in the generic quantized-message models.  The gap-junction model continues tracking at all~$\omega$.  Sweeping the stimulus strength reveals a hysteresis loop in the chemical-synapse model (Fig. 4(C) lower panel), paralleling the hysteresis observed in the generic models.  The bistability arises because the LIF transfer function, an emergent sigmoid shaped by input noise (Eq.~\ref{eq:fIcurve}), combined with slow NMDA-mediated positive feedback, admits two stable self-consistent firing rates (Eq.~\ref{eq:selfconsist}).  Gap-junction currents vanish at consensus voltages and therefore cannot sustain such positive feedback.

We deliberately omit inhibitory synapses from the model~\cite{wang2002probabilistic}.  In cortical decision circuits, mutual inhibition between competing neural groups provides an active mechanism to override lock-in~\cite{wang2002probabilistic}, and its absence here demonstrates that lock-in is an intrinsic property of excitatory chemical transmission alone.

\bibliographystyle{plain}
\bibliography{ColletiveDecision}

\end{document}